\documentstyle[preprint,aps]{revtex}
\renewcommand{\mathbf}[1]{{\bf#1}}
  
\begin{document}
\draft
\title{CLASSICAL ELECTROMAGNETIC INTERACTION OF A POINT CHARGE AND A MAGNETIC MOMENT:CONSIDERATIONS RELATED TO THE AHARONOV-BOHM PHASE SHIFT}
\author{Timothy H. Boyer}
\address{Department of Physics, City College of the City University of New York, New York, NY 10031}
\date{\today}
\maketitle

\begin{abstract}
     A fundamentally new understanding of the classical electromagnetic interaction of a point charge and a magnetic dipole moment through order $v^2/c^2$ is suggested.  This relativistic analysis connects together hidden momentum in magnets, Solem's
 strange polarization of the classical hydrogen atom, and the Aharonov-Bohm phase shift.  First we review the predictions following from the traditional particle-on-a-frictionless-rigid-ring model for a magnetic moment.  This model, which is not rela
tivistic to order $v^2/c^2$, does reveal a connection between the electric field of the point charge and hidden momentum in the magnetic moment; however, the electric field back at the point charge due to the Faraday-induced changing magnetic moment
is of order $1/c^4$ and hence is negligible in a $1/c^2$ analysis.  Next we use a relativistic magnetic moment model consisting of many superimposed classical hydrogen atoms (and anti-atoms) interacting through the Darwin Lagrangian with an external
charge but not with each other.  The analysis of Solem regarding the strange polarization of the classical hydrogen atom is seen to give a fundamentally different mechanism for the electric field of the passing charge to change the magnetic moment.
The changing magnetic moment leads to an electric force back at the point charge which i)is of order $1/c^2$, ii)depends upon the magnetic dipole moment, changing sign with the dipole moment, iii) is odd in the charge $q$ of the passing charge, and i
v) reverses sign for charges passing on opposite sides of the magnetic moment.  Using the insight gained from this relativistic model and the analogy of a point charge outside a conductor, we suggest that a realistic multi-particle magnetic moment in
volves a changing magnetic moment which keeps the electromagnetic field momentum constant.  This means also that the magnetic moment does not allow a significant shift in its internal center of energy.  This criterion also implies that the Lorentz fo
rces on the charged particle and on the point charge are equal and opposite and that the center of energy of each moves according to Newton's second law $\mathbf{F}=M\mathbf{a}$ where $\mathbf{F}$ is exactly the Lorentz force.  Finally, we note that
the results and suggestion given here are precisely what are needed to explain both the Aharonov-Bohm phase shift and the Aharonov-Casher phase shift as arising from classical electromagnetic forces.  Such an explanation reinstates the traditional se
miclassical connection between classical and quantum phenomena for magnetic moment systems.
\end{abstract}

\noindent{\bf 1. INTRODUCTION}

\noindent{\bf a) A New Ingredient Missing from Previous Work}

           The classical electromagnetic interaction of a point charge and a magnetic dipole moment is a poorly understood aspect of classical physics.  However, our understanding of this interaction is important in interpreting several famous experi
ments and is influential in assessing the connections between classical and quantum physics.  Indeed, it turns out that one of the crucial ingredients for understanding this interaction has been missing from both the instructional and research litera
ture.  Although a role for the electrostatic field of the passing charge has been explored in connection with both electromagnetic field momentum and hidden particle momentum in magnets,[1] the role of the electric field in changing the magnetic mome
nt and so in providing an electric force back on the passing charge has been completely missed.  In this article we will discuss the interaction of a point charge and a magnetic dipole moment, noting the inadequacy of previous models, emphasizing the
 previously unrecognized aspects in a relativistic model, and suggesting that the realistic interaction of a point charge and a magnet reinstates both Newton's second and third laws involving the Lorentz force.

\noindent{\bf b) Importance of Understanding the Interaction}

         The interaction of a magnetic moment and a point charge seems like such a basic element within classical physics that one would expect it to be discussed in the electromagnetism textbooks.  However, this is not the case.  Although many texts
 discuss the magnetic interaction between current loops (and by extension between magnetic dipole moments), there seems to be no discussion of the subject of this article.  However, the teaching and research literature is full of discussions of  and
paradoxes involving magnetic moments and point charges.  The Shockly-James paradox[2,3] appears in the research literature while controversy regarding magnetic moments and electromagnetic fields is a staple of the American Journal of Physics.[4]  Ind
eed, in trying to understand the interaction of a magnetic moment and a point charge, various authors have introduced new equations of motion for magnetic moments which depart from Newton's second law involving the Lorentz force.[5]  The interaction
of a magnetic moment and a point charge is at the heart of the experimentally observed Aharonov-Bohm[6,7] and Aharonov-Casher[8,9] phase shifts.  Because of our poor understanding of the classical electromagnetic interactions involved, these experime
nts have been interpreted as revealing a new quantum topological effect representing a new departure between classical and quantum theories of physics.  Thus our understanding of the classical electromagnetic interaction of a magnetic moment and a po
int charge influences our interpretation both of experiments and of fundamental aspects of physical theory.

\noindent{\bf c) Outline of the Discussion}

     This article is organized so as to begin with familiar aspects of magnetic moments and then to include the unexpected elements.  Thus we begin with two current loops and note the requirement that the currents (and hence the magnetic moments) are
 forced to change if there are no batteries in the circuits when the circuits are displaced relative to each other.  Next we turn to the interaction of a point charge and a magnetic moment, discussing the basic electromagnetic forces and pointing out
 that the forces depend upon any changes in the magnetic moment.  We then look at two specific models in order to determine the crucial magnetic moment change.  First we treat the traditional, not-relativistic model of a charge sliding on a frictionl
ess rigid ring, and we conclude that there is no significant change in the magnetic moment and therefore no significant electric field arising due to the magnetic moment.  Second we discuss a relativistic (to order $1/c^2$) model based upon a hydroge
n-atom magnetic moment described by the Darwin Lagrangian.  In this case, Solem's[10 ] observation regarding the "strange polarization of the classical atom" leads to an entirely unanticipated change in the magnetic moment, which leads to significant
 electric fields acting on the passing charged particle.  We then look at the relativistic motion of the center of mass of the magnetic moment and note that this depends crucially upon the change of the magnetic moment.  Working by analogy with the b
ehavior of a point charge outside a conductor and our results from the relativistic magnetic moment model, we suggest that the magnetic moment will change in just such a way as to prevent energy flow within the magnetic moment, keep a constant electr
omagnetic linear momentum for the electromagnetic field, cause the traditional Newton's second law for the motion of the magnetic moment connected to the Lorentz force, and maintain the validity of Newton's third law for the Lorentz forces between th
e point charge and the magnetic moment.  Finally, we note that the proposed behavior of the magnetic moment would give a classical force-based explanation of the Aharonov-Bohm and Aharonov-Casher phase shifts.  The proposed behavior would eliminate t
he quantum topological Aharonov-Bohm effect as the explanation of these phase shifts, and would restore the traditional connection between classical and quantum physics for magnetic moment systems.

\noindent{\bf 2.  FAMILIAR PRELIMINARY: THE INTERACTION OF TWO MAGNETIC MOMENTS}

\noindent{\bf a) Magnetic Moments in Electromagnetism Texts}

     Although the interaction of a point charge with a magnetic moment does not appear in the textbooks, the magnetic moment interaction which does appear is that of two current loops with steady currents.  In the limit of small current loops, we can
 think of the interaction as involving two magnetic moments.  In this case the results are familiar and reassuring.  The magnetic Lorentz forces of one circuit on the other satify Newton's third
law, giving forces of equal magnitude and opposite direction.[11]  In the limit of  current loops which are small compared to their separation, we can write the magnetic forces as [12]
\begin{eqnarray*}
\mathbf{F}_{on 1}=\nabla_1[\mathbf{m}_1 \cdot\mathbf{B}_2(\mathbf{r}_1,t)]=\nabla_1[\mathbf{m}_1 \cdot \nabla_1 \times \mathbf{A}_2(\mathbf{r}_1,t)]
\end{eqnarray*}
\begin{eqnarray*}
=\nabla_1\left(\mathbf{m}_1 \cdot \left\{\nabla_1\times\left[\frac{\mathbf{m}_2\times(\mathbf{r}_1-\mathbf{r}_2)}{|\mathbf{r}_1-\mathbf{r}_2|^3}\right]\right\}\right)
=-\nabla_2\left(\mathbf{m}_2 \cdot \left\{\nabla_2\times\left[\frac{\mathbf{m}_1\times(\mathbf{r}_2-\mathbf{r}_1)}{|\mathbf{r}_2-\mathbf{r}_1|^3}\right]\right\}\right)
\end{eqnarray*}
\begin{equation}
=-\nabla_2[\mathbf{m}_2 \cdot \nabla_2 \times \mathbf{A}_1(\mathbf{r}_2,t)]=-\nabla_2[\mathbf{m}_2 \cdot\mathbf{B}_1(\mathbf{r}_2,t)]=-\mathbf{F}_{on 2}
\end{equation}
Here there is no electrostatic field present, and  hence no linear momentum in the electromagnetic field.  In this case we expect the validity of Newton's second law for the motion of the center of energy of each magnetic moment

\noindent{\bf b) Energy Considerations and Changing Magnetic Moments}

          Although the magnetic forces between two magnetic dipole moments reassuringly satisfy Newton's third law, just as do the forces in electrostatics, the energy considerations are emphatically and disconcertingly different from the electrostat
ic situation.  Here we are reminded that magnetic moments involve hidden aspects which are unfamiliar from our experience with electrostatics.  If one circuit of the pair considered above is rigidly and quasistatically displaced, then there is net me
chanical work done associated with the net Lorentz force on the circuit; however, there is, of necessity, other work involved.  There is work associated with the currents of the magnetic moments moving in the emfs induced by the relative movement of
the circuits.  If any batteries in the circuits hold the currents constant, then the work done by each battery is equal in magnitude but opposite in sign to the mechanical work done in displacing one of the circuits.[13]  In the case of small circuit
s which can be regarded as magnetic moments, we have
\begin{equation}
\delta W_{battery} = I\delta\Phi=-\delta W_{mechanical}
\end{equation}
When there are no batteries present, then the internal energy of each magnetic moment will change because of the work done by the emf acting on the currents. This energy change is a hidden aspect of a magnetic moment.

      For toy magnets we use $\mathbf{F}=M\mathbf{a}$ where $\mathbf{F}$ is calculated as the Lorentz force between current loops in (1) above and $\mathbf{a}$ is the acceleration of the center of energy of the magnet.  When the magnets are modeled a
s current loops, then the energies $U$ of the magnets change, but these energy changes are of order $1/c^2$, giving mass changes $M=U/c^2$ of order $1/c^4$, and so do not affect the acceleration through order $1/c^2$.

\noindent{\bf 3. PRELIMINARY DISCUSSION OF THE INTERACTION OF A POINT CHARGE AND A MAGNETIC MOMENT}

\noindent{\bf a) The Lorentz Forces and Electromagnetic Field Momentum}

          In the analysis of the present article, we are interested in relativistic effects through order $1/c^2$.  This corresponds to the quasistatic approximation to classical electromagnetic theory.  In this quasistatic approximation the fields d
ue to a magnetic moment at a fixed position $\mathbf{r}_m$ can  be described in terms of a vanishing scalar potential $\Phi_m(\mathbf{r},t)$ and a magnetic vector potential $\mathbf{A}_m(\mathbf{r},t)$ of a magnetic moment,[14]
\begin{equation}
\Phi_m(\mathbf{r},t)=0,
\mathbf{A}_m(\mathbf{r},t)=\mathbf{m}(t)\times\frac{(\mathbf{r}-\mathbf{r}_m)}{|\mathbf{r}-\mathbf{r}_m|^3}
\end{equation}
with the associated electric and magnetic fields
\begin{equation}
\mathbf{E}_m(\mathbf{r},t)=-\frac{1}{c}\frac{\partial\mathbf{A}_m(\mathbf{r},t)}{\partial t}=-\frac{1}{c}\frac{d\mathbf{m}}{dt}\times\frac{\mathbf{r}-\mathbf{r}_m}{|\mathbf{r}-\mathbf{r}_m|^3}
\end{equation}
\begin{equation}
\mathbf{B}_m(\mathbf{r},t)=\nabla\times\mathbf{A}_m(\mathbf{r},t)=
\frac{1}{|\mathbf{r}-\mathbf{r}_m|^3}\left(\frac{3[\mathbf{m}\cdot(\mathbf{r}-\mathbf{r}_m)](\mathbf{r}-\mathbf{r}_m)}{|\mathbf{r}-\mathbf{r}_m|^2}-\mathbf{m}\right)
\end{equation}

     In this same quasistatic approximation, the fields due to a point charge can be described in terms of scalar and vector potentials [15]
\begin{equation}
\Phi_q(\mathbf{r},t)=\frac{q}{|\mathbf{r}-\mathbf{r}_q|}  , \mathbf{A}_q(\mathbf{r},t)=\frac{q}{2c|\mathbf{r}-\mathbf{r}_q|}\left(\mathbf{v}_q+\frac{[\mathbf{v}_q\cdot(\mathbf{r}-\mathbf{r}_q)](\mathbf{r}-\mathbf{r}_q)}{|\mathbf{r}-\mathbf{r}_q|^2}\right)
\end{equation}
with the associated electric and magnetic fields
\begin{eqnarray*}
\mathbf{E}_q(\mathbf{r},t)=-\nabla\Phi_q-\frac{1}{c}\frac{\partial\mathbf{A}_q}{\partial t}
\end{eqnarray*}
\begin{equation}
=q\frac{(\mathbf{r}-\mathbf{r}_q)}{|\mathbf{r}-\mathbf{r}_q|^3}\left(1+\frac{1}{2}\frac{v^2_q}{c^2}-\frac{3}{2}\frac{[\mathbf{v}_q\cdot(\mathbf{r}-\mathbf{r}_q)]^2}{|\mathbf{r}-\mathbf{r}_q|^2}\right)-\frac{q}{2c|\mathbf{r}-\mathbf{r}_q|}\left(\mathbf{a}_q+\frac{[\mathbf{a}_q\cdot(\mathbf{r}-\mathbf{r}_q)](\mathbf{r}-\mathbf{r}_q)}{|\mathbf{r}-\mathbf{r}_q|^2}\right)
\end{equation}
\begin{equation}
\mathbf{B}_q(\mathbf{r},t)=\frac{q}{c}\mathbf{v}_q\times\frac{(\mathbf{r}-\mathbf{r}_q)}{|\mathbf{r}-\mathbf{r}_q|^3}
\end{equation}

     Thus when a point charge is located near a magnetic moment, these electromagnetic distributions interact through their electromagnetic fields producing a Lorentz force on the magnetic moment $\mathbf{m}$ given by[16]
\begin{equation}
\mathbf{F}_{on m}=\nabla_m[\mathbf{m}\cdot\mathbf{B}_q(\mathbf{r}_m,t)]=\nabla_m\left[\mathbf{m}\cdot\left(\frac{q}{c}\mathbf{v}_q\times\frac{(\mathbf{r}_m-\mathbf{r}_q)}{|\mathbf{r}_m-\mathbf{r}_q|^3}\right)\right]=-\nabla_q\left[\frac{q}{c}\mathbf{
v}_q\cdot\mathbf{A}_m(\mathbf{r}_q,t)\right]
\end{equation}
and a Lorentz force on the passing charge $q$ given by
\begin{eqnarray*}
\mathbf{F}_{on q}=q\mathbf{E}_m(\mathbf{r}_q,t)+(q/c)\mathbf{v}_q\times\mathbf{B}_m(\mathbf{r}_q,t)
\end{eqnarray*}
\begin{eqnarray*}
=-\frac{q}{c}\frac{d\mathbf{m}}{dt}\times\left(\frac{\mathbf{r}_q-\mathbf{r}_m)}{|\mathbf{r}_q-\mathbf{r}_m|^3}\right)+\frac{q}{c}\mathbf{v}_q\times\left[\nabla_q\times\mathbf{A}_m(\mathbf{r}_q,t)\right]
\end{eqnarray*}
\begin{equation}
=\frac{1}{c}\frac{d\mathbf{m}}{dt}\times\mathbf{E}_q(\mathbf{r}_m,t)+\frac{q}{c}\mathbf{v}_q\times[\nabla_q\times\mathbf{A}_m(\mathbf{r}_q,t)]
\end{equation}
In general these forces are not equal in magnitude and opposite in direction, in contrast to the situation which held in the case of two interacting magnetic moments.  Now if the forces do not satisfy Newton's third law, then conservation of linear m
omentum requires the presence of some additional momentum; in this case the additional linear momentum is located in the electromagnetic field.  Indeed, the presence of the electric field of the point charge q leads to an electromagnetic field linear
 momentum[17]
\begin{equation}
\mathbf{P}_{em}=\frac{1}{4\pi c}\int d^3r\mathbf{E}_q\times\mathbf{B}_m=\frac{q}{c}\mathbf{A}_m(\mathbf{r}_q,t)=\frac{q}{c}\frac{\mathbf{m}\times(\mathbf{r}_q-\mathbf{r}_m)}{|\mathbf{r}_q-\mathbf{r}_m|^3}
\end{equation}
Now if we sum the Lorentz forces, we find from Eqs. (9) and (10)
\begin{eqnarray*}
\mathbf{F}_{on m}+\mathbf{F}_{on q}=-\nabla_q[\frac{q}{c}\mathbf{v}_q\cdot\mathbf{A}_m(\mathbf{r}_q,t)]
+\left(\frac{1}{c}\frac{d\mathbf{m}}{dt}\times\mathbf{E}_q(\mathbf{r}_m,t)+\frac{q}{c}\mathbf{v}_q\times[\nabla_q\times\mathbf{A}_m(\mathbf{r}_q,t)]\right)
\end{eqnarray*}
\begin{equation}
=-\frac{q}{c}(\mathbf{v}_q\cdot\nabla_q)\mathbf{A}_m(\mathbf{r}_q,t)+\frac{1}{c}\frac{d\mathbf{m}}{dt}\times\mathbf{E}_q(\mathbf{r}_m,t)=-\frac{d}{dt}\left(\frac{q}{c}\frac{\mathbf{m}\times(\mathbf{r}_q-\mathbf{r}_m)}{|\mathbf{r}_q-\mathbf{r}_m|^3}\right)=-\frac{d}{dt}\mathbf{P}_{em}
\end{equation}
where the time-dependence appears in $\mathbf{m}(t)$ and $\mathbf{r}_q(t)$ in this inertial frame.  This indeed provides conservation of the total linear momentum of the system since Eq. (12) holds as a vector identity.  We notice that the two Lorent
z forces satisfy Newton's third law if and only if the linear momentum in the electromagnetic field does not change in time
\begin{equation}
\frac{d}{dt}\mathbf{P}_{em}=0
\end{equation}
However, the constancy of $\mathbf{P}_{em}$ depends upon the time rate of change of the magnetic moment, $d\mathbf{m}/dt$.  Later we will suggest that the magnetic moment actually changes in just such a way as to keep $\mathbf{P}_{em}$ constant so th
at the Lorentz forces between the point charge and the magnetic dipole moment indeed satisfy Newton's third law.

\noindent{\bf b) Constant Motion of the Center of Energy}

     In addition to conservation of linear momentum, relativistic theories also require uniform motion of the system center of energy.[1]  Now within nonrelativistic physics, the statement of conservation of linear momentum is the same as the stateme
nt of uniform motion of the center of mass; within relativistic physics, the statements are different.  Thus if we identify the mass $M_m$ of the magnetic moment as its relativistic energy $U_m$ divided by $c^2$, then (since the magnetic field energy
 is already of order $1/c^2$) we require[18]
\begin{equation}
M_m\mathbf{a}_m+M_q\mathbf{a}_q =0
\end{equation}
However, this equation is not easily reconciled with equation (12) for forces satisfying Newton's second law, unless the condition of Eq.(13) actually holds.

     Indeed, because of these relativisitic complications, Aharonov, Pearl, and Vaidman,[19] along with others [20], have introduced a equation of motion for the magnetic moment which is different from Newton's $\mathbf{F}=M\mathbf{a}$ where $\mathbf
{F}$ is the Lorentz force.  In order to gain an understanding of these matters, we must determine $d\mathbf{m}/dt$.  This is a crucial element in the analysis.  However, in order to do this, it is not sufficient to work from the prescribed charge and
 current distributions for magnetic moments which appear in the text books.  Rather we must turn to specific models for the behavior of the magnetic moment.  In this article, we will look at two different models for magnetic moment behavior which lea
d to vastly different results for $d\mathbf{m}/dt$.

\noindent{\bf 4. THE PARTICLE-ON-A-RIGID-RING MODEL FOR A MAGNETIC MOMENT}

\noindent{\bf a) The Model}

          The model for a magnetic dipole moment which appears frequently in both the teaching and and research literature is that of one or more charged particles $e_b$ sliding around a small, rigid, frictionless, circular ring.  The masses $m_b$ of
 the moving charges are taken as small while the mass of the circular ring is large, and the circular ring has at its center a  charge which is opposite from that of the moving charges.  The magnetic dipole moment is given as
\begin{equation}
\mathbf{m}=\sum_b\frac{e_b}{2c}\mathbf{r}_b\times\mathbf{v}_b=\sum_b\mathbf{k}\frac{e_b}{2c}R^2\omega_b
\end{equation}
where the sum is over all the moving charges $e_b$ moving with velocity $\mathbf{v}_b$, and here we have chosen the ring to lie in the plane $z=0$ and the particle speeds $v_b=R\omega_b$.  The use of a single charge $e_b$ makes the calculations easie
st.  By appropriately stacking these circular magnetic-moment loops, we can form a solenoid or a toroid.

      The electromagnetic fields caused by the model and the Lorentz forces experienced by the model, when averaged over the circular motion, are taken as those of the magnetic moment.  The motion of the charge $e_b$ is circular
\begin{equation}
\mathbf{r}_b=R[\mathbf{i}cos(\theta_b)+\mathbf{j}sin(\theta_b)]
\end{equation}
\begin{equation}
\mathbf{v}_b=R\omega_b[-\mathbf{i}sin(\theta_b)+\mathbf{j}cos(\theta_b)]
\end{equation}
\begin{equation}
\mathbf{a}_b=-R\omega_b^2[\mathbf{i}cos(\theta_b)+\mathbf{j}sin(\theta_b)]
+R\alpha_b[-\mathbf{i}sin(\theta_b)+\mathbf{j}cos(\theta_b)]
\end{equation}
where we have taken the center of the circular loop as the origin of coordinates and have written $\omega_b=d\theta_b/dt$, $\alpha_b=d\omega_b/dt$.  For the unperturbed motion, $\alpha_b=0$.

\noindent{\bf b) The Lorentz Forces}

         It is easy to calculate the average Lorentz forces between the model magnetic moment $\mathbf{m}$ and a passing charge $q$.  To obtain the force on the charge q, we evaluate the average electric and magnetic fields due to the magnetic moment
.  The fields follow from the average electromagnetic potentials.  The magnetic vector potential $\mathbf{A}_m(\mathbf{r},t)$ produced by the circulating charge $e_b$ at a field point $\mathbf{r}$, when averaged over the steady-state motion (16) in t
he limit $R<<r$ is just[21]
\begin{eqnarray*}
\mathbf{A}_m(\mathbf{r},t)=\left<\frac{e_b}{2c|\mathbf{r}-\mathbf{r}_b|}\left(  \mathbf{v}_b+\frac{[\mathbf{v}_b\cdot(\mathbf{r}-\mathbf{r}_b)](\mathbf{r}-\mathbf{r}_b)}{|\mathbf{r}-\mathbf{r}_b|^2}\right)\right>
\end{eqnarray*}
\begin{eqnarray*}
=\frac{e_b}{2cr}\left<\left(1+\frac{\mathbf{r}\cdot\mathbf{r}_b}{r^2}  \right)\left[\mathbf{v}_b+\frac{[\mathbf{v}_b\cdot(\mathbf{r}-\mathbf{r}_b)](\mathbf{r}-\mathbf{r}_b)}{r^2}\left(1+\frac{2\mathbf{r}\cdot\mathbf{r}_b}{r^2}  \right)\right]\right>
\end{eqnarray*}
\begin{eqnarray*}
=\frac{e_b}{2cr^3}\left<\mathbf{v}_b(\mathbf{r}_b\cdot\mathbf{r})-\mathbf{r}_b(\mathbf{v}_b\cdot\mathbf{r})\right>
\end{eqnarray*}
\begin{equation}
=\frac{e_b}{2c}<\mathbf{r}_b\times\mathbf{v}_b>\times\frac{\mathbf{r}}{r^3}=\frac{\mathbf{m}\times\mathbf{r}}{r^3}
\end{equation}
just as in Eq.(3) for the vector potential of a magnetic moment located at the origin of coordinates.  The magnetic field $\mathbf{B}_m(\mathbf{r},t)$ of the model follows as in Eq.(8) corresponding to the magnetic field of a magnetic dipole $\mathbf
{m}$ centered on the origin of coordinates.  The electric potential, when averaged over the steady-state motion in the limit $R<<r$, gives
\begin{eqnarray*}
\Phi_m(\mathbf{r},t)=\left<\frac{-e_b}{r}+\frac{e_b}{|\mathbf{r}-\mathbf{r}_b|}\right>
\end{eqnarray*}
\begin{equation}
=\frac{e_b}{r}\left<\frac{\mathbf{r}\cdot\mathbf{r}_b}{r^2}-\frac{r^2_b}{2r^2}+\frac{3}{2}\left(\frac{\mathbf{r}\cdot\mathbf{r}_b}{r^2}\right)^2  \right>=\frac{\mathbf{r}\cdot\mathbf{Q}\cdot\mathbf{r}}{2r^5}
\end{equation}
where $\mathbf{Q}$ is the electric quadrupole tensor
\begin{equation}
\mathbf{Q}=e_b<3\mathbf{r}_b\mathbf{r}_b-\mathbf{I}r^2_b>
\end{equation}
We note that the overall distribution is neutral and the rotating electric dipole vanishes when averaged over the steady-state motion.  There remains the electric field from the average electric quadrupole moment, but this is not of interest since it
 reverses sign when both the charge $e_b$ and the angular velocity $\omega_b=v_b/R$ are reversed in sign so as to maintain the sign of the magnetic dipole moment.  Thus, ignoring the electric quadrupole term, the electric field of the model gives acc
ordingly
\begin{eqnarray*}
\mathbf{E}_m(\mathbf{r},t)=-\nabla\Phi_m(\mathbf{r},t)-\frac{1}{c}\frac{\partial\mathbf{A}_m(\mathbf{r},t)}{\partial t}
\end{eqnarray*}
\begin{equation}
=-\frac{1}{c}\frac{\partial\mathbf{A}_m(\mathbf{r},t)}{\partial t}
=-\frac{1}{c}\frac{d\mathbf{m}}{dt}\times\frac{\mathbf{r}}{r^3}
\end{equation}
The Lorentz force $\mathbf{F}_{onq}$ on the passing charge $q$ due to the model magnetic moment is exactly as in Eq.(10)

     The Lorentz force experienced by the magnetic moment is found from the Lorentz force on the circulating charge when averaged over the orbit for $R<<r$,
\begin{eqnarray*}
\mathbf{F}_{on m}=\left<-e_b\mathbf{E}_q(\mathbf{r}_{b'})_{\mathbf{r}_{b'}=0},t)+e_b\mathbf{E}_q(\mathbf{r}_b,t)+e_b\frac{\mathbf{v}_b}{c}\times\mathbf{B}_q(\mathbf{r}_b,t) \right>=
\end{eqnarray*}
\begin{equation}
=\left<e_b\frac{1}{2}(\mathbf{r}_b\cdot\nabla)^2\mathbf{E}_q(\mathbf{r},t)_{\mathbf{r}=0}+e_b\frac{\mathbf{v}_b}{c}\times[\mathbf{B}_q(\mathbf{r},t)_{\mathbf{r}=0}+\left\{(\mathbf{r}_b\cdot\nabla)\mathbf{B}_q(\mathbf{r})\right\}_{\mathbf{r}=0}]   \right>
\end{equation}
Now the first term on the second line of Eq.(23)can be neglected as the force on the electrical quadrupole moment (which vanishes under the magnetic-moment preserving sign-reversal of both $e_b$ and $\omega$), and the term involving $e_b\mathbf{v}_b$
 vanishes when averaged over the steady-state motion.  The term involving $e_b\mathbf{v}_b\mathbf{r}_b$ can be transformed using tenor notation when we note that the average $<x_ix_j>$ is time-independent for a steady-state motion and so that $d<x_ix
_j>/dt=<x_iv_j+v_ix_j>=0$, and also note that $\nabla\cdot\mathbf{B}=\partial_iB_i=0$.  Thus we have
\begin{eqnarray*}
<\{\mathbf{v}\times[(\mathbf{r}\cdot\nabla)\mathbf{B}]\}_i>=<\epsilon_{ijk}v_jx_l\partial_lB_k>
\end{eqnarray*}
\begin{eqnarray*}
=<(1/2)\epsilon_{ijk}(v_jx_l-v_lx_j)\partial_lB_k>=<(1/2)\epsilon_{ijk}\epsilon_{mjl}\epsilon_{mpq}v_px_q\partial_lB_k>
\end{eqnarray*}
\begin{eqnarray*}
=<(1/2)(\delta_{kl}\delta_{im}-\delta_{km}\delta_{il})\epsilon_{mpq}v_px_q\partial_lB_k>
\end{eqnarray*}
\begin{eqnarray*}
=<(1/2)\epsilon_{ipq}v_px_q\partial_kB_k>-<(1/2)\epsilon_{kpq}v_px_q\partial_iB_k>
\end{eqnarray*}
\begin{equation}
=<(1/2)\epsilon_{kpq}x_pv_q\partial_iB_k>=<\{\nabla[(1/2)(\mathbf{r}\times\mathbf{v})\cdot\mathbf{B}]\}_i>
\end{equation}
where $\nabla$ acts on the argument of $\mathbf{B}$ and not on the $(\mathbf{r}\times\mathbf{v})$ which should actually have b-subscripts.  Accordingly, we have our required result
\begin{equation}
\mathbf{F}_{onm}=\nabla[\mathbf{m}\cdot\mathbf{B}_q(\mathbf{r},t)]_{r=0}
\end{equation}
where the magnetic moment is as given in Eq.(15).   This agrees with the force given in Eq.(9).

     The linear momentum in the electromagnetic field can be obtained from the usual expression involving point charges,[22] again averaging over the steady-state orbit,
\begin{eqnarray*}
\mathbf{P}_{em}=\frac{q(-e_b)}{2cr_q}\left(\mathbf{v}_q+\frac{(\mathbf{v}_q\cdot\mathbf{r}_q)\mathbf{r}_q}{r^2_q}   \right)+\frac{qe_b}{2c|\mathbf{r}_q-\mathbf{r}_b|}\left(\mathbf{v}_b+\mathbf{v}_q+\frac{[(\mathbf{v}_b+\mathbf{v}_q)\cdot(\mathbf{r}_b
-\mathbf{r}_q)](\mathbf{r}_b-\mathbf{r}_q)}{|\mathbf{r}_b-\mathbf{r}_q|^2}   \right)
\end{eqnarray*}
\begin{eqnarray*}
=\frac{qe_b}{2cr_q}\left[-\mathbf{v}_q-\frac{(\mathbf{v}_q\cdot\mathbf{r}_q)\mathbf{r}_q}{r^2_q}\right]
\end{eqnarray*}
\begin{equation}
+\frac{qe_b}{2cr_q}\left[\left(1+\frac{\mathbf{r}_q\cdot\mathbf{r}_b}{r^2_q}  \right)\left\{\mathbf{v}_b+\mathbf{v}_q+\frac{[(\mathbf{v}_q+\mathbf{v}_b)\cdot(\mathbf{r}_b-\mathbf{r}_q)](\mathbf{r}_b-\mathbf{r}_q)}{r^2_q}\left(  1+\frac{2\mathbf{r}_q\
cdot\mathbf{r}_b}{r^2_q}\right)  \right\} \right]
\end{equation}
where we assume $r_b<<r_q$ corresponding to a small magnetic moment, and retain only first-order terms in $r_b/r_q$.  Then multiplying out Eq.(26), removing the canceling terms, and averaging over the steady-state motion,
\begin{eqnarray*}
<\mathbf{P}_{em}>=\frac{qe_b}{2cr_q}\left<+\mathbf{v}_b+\frac{(\mathbf{r}_q\cdot\mathbf{r}_b)\mathbf{v}_b}{r^2_q}+\frac{(\mathbf{r}_q\cdot\mathbf{r}_b)\mathbf{v}_q}{r^2_q}+
+\frac{(\mathbf{v}_q\cdot\mathbf{r}_b)(-\mathbf{r}_q)}{r^2_q}+\frac{(\mathbf{v}_q\cdot\mathbf{r}_q)(-\mathbf{r}_b)}{r^2_q}\right>
\end{eqnarray*}
\begin{eqnarray*}
+\frac{qe_b}{2cr_q}\left<\frac{(\mathbf{v}_b\cdot\mathbf{r}_q)\mathbf{r}_q}{r^2_q}
+\frac{(\mathbf{v}_b\cdot\mathbf{r}_b)(-\mathbf{r}_q)}{r^2_q}+\frac{(\mathbf{v}_b\cdot\mathbf{r}_q)(-\mathbf{r}_b)}{r^2_q}+3\frac{(\mathbf{r}_q\cdot\mathbf{r}_b)}{r^2_q}\frac{(\mathbf{v}_q\cdot\mathbf{r}_q)\mathbf{r}_q}{r^2_q}+3\frac{(\mathbf{r}_q  \cdot\mathbf{r}_b)}{r^2_q}\frac{(\mathbf{v}_b\cdot\mathbf{r}_q)\mathbf{r}_q}{r^2_q}  \right>
\end{eqnarray*}
\begin{equation}
=\frac{qe_b}{2cr^3_q}\left<(\mathbf{r}_q\cdot\mathbf{r}_b)\mathbf{v}_b-(\mathbf{r}_q\cdot\mathbf{v}_b)\mathbf{r}_b  \right>
=\frac{e_b}{2c}\left<\mathbf{r}_b\times\mathbf{v}_b \right>\times\frac{q\mathbf{r}_q}{r^3_q} = -\mathbf{m}\times\mathbf{E}_q(0)
\end{equation}
Here we have used the fact that $<x_{bi}>=0$, $<v_{bi}>=0$, and  $<x_{bi}v_{bi}>=(d/dt)<\frac{1}{2}x_{bi}^2>=0$ (no sum over $i$) for the steady-state motion.  This result for the average momentum in the electromagnetic field agrees with the general
result given earlier in Eq.(11).

\noindent{\bf c) The Electric Field Back at the Point Charge in this Model}

      We have just seen that our particle-on-a-rigid-ring model for a magnetic moment indeed satisfies the requirements for the electromagnetic fields and forces of a general magnetic moment given in Section 3.  Thus it is of interest to determine wh
at this model gives for the time rate of change of the magnetic moment $d\mathbf{m}/dt$ due to a passing charge $q$, and so to determine the electric field due to the magnetic moment which acts on the passing charge $q$.  The magnetic moment $\mathbf
{m}=(e_b/2c)<\mathbf{r}_b\times\mathbf{v}_b>$ is proportional to the nonrelativistic particle angular momentum $\mathbf{L}$ with the familiar factor $e_b/2M_bc$,
\begin{equation}
\mathbf{m}=\frac{e_b}{2c}<\mathbf{r}_b\times\mathbf{v}_b>=\frac{e_b}{2M_bc}<\mathbf{r}_b\times(M_b\mathbf{v}_b)>=\frac{e_b}{2M_bc}\mathbf{L}
\end{equation}
Thus the rate of change of the magnetic moment $d\mathbf{m}/dt$ is proportional to the rate of change of the particle angular momentum $d\mathbf{L}/dt$ and proportional to the torque.

     In this model, the electrostatic part of the electric field $\mathbf{E}_q$ of the passing charge is essentially uniform across the small magnetic moment; the electrostatic part causes the charge $e_b$ of the magnetic moment to accelerate symetri
cally on opposite sides of the frictionless rigid ring and provides no average torque.  Thus the electrostatic field does not change the magnitude of the magnetic moment $\mathbf{m}$.  However, the magnetic field of the passing charge $q$ does induce
 an emf in the magnetic moment orbit which will lead to a change $d\mathbf{m}/dt$ in the magnetic moment $\mathbf{m}$ and hence to an electric field and force back at the passing charge $q$.  (The part of the electric field corresponding to the induc
ed emf appears in the $v^2_q/c^2$ terms in the electric field expression (7).)   The induced emf in the rigid current loop in the xy-plane is
\begin{equation}
emf=-\frac{1}{c}\frac{d\Phi}{dt}=-\frac{1}{c}\frac{d}{dt}\left(\pi R^2\mathbf{k}\cdot\frac{q\mathbf{v}_q\times(-\mathbf{r}_q)}{cr^3_q}   \right)=\frac{\mathbf{v}_q}{c}\cdot\nabla_q\left(\pi R^2\frac{q\mathbf{k}\cdot(\mathbf{v}_q\times\mathbf{r}_q}{cr
^3_q}\right)
\end{equation}
The average tangential acceleration of the charge $e_b$ is given by the nonrelativistic expression since the force associated with the emf is already of order $1/c^2$,
\begin{equation}
\mathbf{a}=\frac{e_b(emf)}{M_b2\pi R}
\end{equation}
The time rate of change of the magnetic moment is therefore
\begin{equation}
\frac{d\mathbf{m}}{dt}=\frac{e_b}{2c}\mathbf{r}_b\times\mathbf{a}_b = \frac{e_b}{2c}\mathbf{r}_b\times\left\{\frac{e_b}{M_b2\pi R}\left[\frac{\mathbf{v}_q}{c}\cdot\nabla_q\left(\pi R^2\mathbf{k}\cdot\frac{q\mathbf{v}_q\times\mathbf{r}_q}{cr^3_q}
\right)\right]\right\}
\end{equation}
The changing magnetic moment causes an average electric field back at the passing charge which produces a force on $q$
\begin{eqnarray*}
<\mathbf{F}^{electric}_{onq}>=q\left(\frac{\mathbf{r}_q}{cr^3_q}\times\frac{d\mathbf{m}}{dt}  \right)
\end{eqnarray*}
\begin{equation}
=q\left(\frac{\mathbf{r}_q}{cr^3_q}\right)\times\left(\frac{e_b}{2c}\mathbf{r}_b\times\left\{\frac{e_b}{M_b2\pi R}\left[\frac{\mathbf{v}_q}{c}\cdot\nabla_q\left(\pi R^2\mathbf{k}\cdot\frac{q\mathbf{v}_q\times\mathbf{r}_q}{cr^3_q}     \right)\right]\right\}\right)
\end{equation}

     We notice that this induced electric field back at the position of the passing charge $q$  i) is of order $e^2q^2$ and therefore does not change sign with q, ii) does not depend upon the velocity of the charge $e_b$ and therefore does not depend
 upon the magnitude or direction of the magnetic moment, and iii) is of order $1/c^4$ and hence is completely negligible compared to the magnetic Lorentz forces of order $1/c^2$.  [23]  Thus according to this model, the force on the passing charge ar
ises solely from the magnetic field due to the magnetic moment.

\noindent{\bf d) The Hidden Momentum}

          Although the rigid-ring model does not suggest the existence of any significant electric force back on the passing charge $q$, it does suggest the possiblity of hidden momentum in the magnetic moment.  This hidden momentum is easy to calcul
ate in the rigid-ring model and now appears in electromagnetism textbooks.[24]  The circular motion of the charge $e_b$ in the electrostatic field of the passing charge involves a periodic change in the nonrelativistic kinetic energy as the charge mo
ves around the rigid ring; energy conservation gives
\begin{equation}
(1/2)M_bv^2_b+e_bV=(1/2)M_bv^2_{b0}
\end{equation}
where $V=q\mathbf{r}_b\cdot\mathbf{r}_q/r^3_q$ is the changing part of the electrostatic potential of the point charge $q$ when evaluated at the position $\mathbf{r}_b=R(\mathbf{i}cos\omega t + \mathbf{j}sin\omega t)$ of the circulating charge $e_b$,
 and $(1/2)M_bv^2_{b0}$ is the kinetic energy of the circulating charge before the arrival of the passing charge $q$.  Clearly the electrostatic potential is as often positive as negative and so the average energy of the circulating charge $e_b$ is s
till the $(1/2)M_bv^2_{b0}$ which held before the arrival of $q$.  However, in this rigid-ring model, the linear momentum has acquired a new average value (a hidden momentum) given by
\begin{eqnarray*}
<\mathbf{p}_b>=<M_b(1+\frac{v^2_b}{2c^2})\mathbf{v}_q>=<M_b(1+\frac{v^2_{b0}}{2c^2})\mathbf{v}_b>+\frac{1}{c^2}<-e_bV\mathbf{v}_b>
\end{eqnarray*}
\begin{eqnarray*}
=<M_b(1+\frac{v^2_{b0}}{2c^2})R\omega(-\mathbf{i}sin\omega t+\mathbf{j}cos\omega t)> +\frac{1}{c^2}\left<\left(\frac{-e_bqR}{r^3_q}(x_qcos\omega t+y_qsin\omega t)R\omega(-\mathbf{i}sin\omega t+\mathbf{j}cos\omega t) \right)\right>
\end{eqnarray*}
\begin{equation}
=0-\frac{e_bqR^2\omega}{c^2r^3_q}(-\mathbf{i}y_q+\mathbf{j}x_q)\frac{1}{2}=-\frac{q}{c}\frac{\mathbf{m}\times\mathbf{r}_q}{r^3_q}=\frac{q}{c}\mathbf{m}\times\mathbf{E}_q(0,t)
\end{equation}
where $\mathbf{E}_q(0,t)$ is the electric field due to the external charge $q$ evaluated at the position of the magnetic moment $\mathbf{r}_m=0$.[25]

\noindent{\bf e) New Proposed Equation of Motion}

     Since we have aready seen that in this rigid-ring model the change in the magnetic moment $d\mathbf{m}/dt$ is negligible (as order $1/c^2$ and leading to electric fields of order $1/c^4$), it has been suggested[19] that part of the magnetic Lore
ntz force (25) acting on the magnetic moment goes simply into changing this "hidden momentum" and thus the center of energy of the magnetic moment accelerates according to
\begin{equation}
M_m\mathbf{a}_m=\mathbf{F}_{on m}-\frac{q}{c}\mathbf{m}\times\frac{d\mathbf{E}(0,t)}{dt}
\end{equation}
This is a strange equation of motion where Newton's second law no longer connects the acceleration of a mass to a force given by the Lorentz force $\mathbf{F}_{onm}$.  However, this equation of motion for a magnetic moment or a magnet has been accept
ed by numbers of physicists.

\noindent{\bf 5. A RELATIVISTIC MODEL FOR A MAGNETIC MOMENT}

\noindent{\bf a) The Model}

     In theoretical physics we have the greatest confidence when our results are model-independent.  In the previous section, we evaluated the interaction of a point charge and a magnetic dipole moment using the charge-on-a-rigid-ring model for the m
agnetic moment.  This model, although thoroughly familiar and easy to work with, is not relativistic to order $v^2/c^2$.  The nonrelativisitic character is immediately betrayed by the rigid quality of the ring.  Furthermore, this model violates the r
elativistic requirement of uniform motion of the center of energy since it involves the continual transfer of energy from one side of the ring to the other without any compensating shift in the system center of energy.  Thus we turn next to a relativ
isitic model for the magnetic moment in order to check whether our previous conclusions are still valid in a different model.

     The Darwin Lagrangian, which is relativistic through order $1/c^2$, is known to represent the classical electromagnetic interaction between charged particles through order $1/c^2$.[26]  Thus for a relativistic magnetic moment model, we will choo
se not a circulating charge $e_b$ on a rigid ring, but rather a charge $e_b$ in a circular Coulomb orbit around a much heavier nucleus of charge $e_{b'}=-e_b$.  This hydrogen-atom system with a circular orbit has a time-average magnetic dipole moment
 which is the same as that of a charge sliding on a rigid, circular, frictionless ring, provided the charge $e_b$, radius $R$, and angular frequency $\omega$ are the same.    Further, we imagine that the magnetic dipole moment consists of a superposi
tion of non-interacting hydrogen atoms (and anti-atoms with opposite angular velocity $\omega$) so as to eliminate all electrostatic multipole moments while maintaining the magnetic dipole moment.  In our model, the external charge $q$ interacts thro
ugh the Darwin Lagrangian with each of the atoms which make up the magnetic moment, but none of the atoms interact with other atoms.

     If the unperturbed orbits are the same, we see immediately that our calculations for the average magnetic Lorentz forces between the magnetic moment and an external charge $q$ are identical to those obtained in the previous section using the rig
id-ring model for a magnetic moment.  The same agreement holds for the electromagnetic field momentum involving the magnetic moment and the point charge.  The calculations of these quantities depended simply upon the stationary nature of the average
currents of the magnetic dipole in the absence of any interaction between the point charge and the magnetic moment.

\noindent{\bf b) New Behavior:Solem's Strange Polarization of the Classical Hydrogen Atom}

      Although the magnetic Lorentz forces and the electromagnetic field momentum depend only upon the average stationary character of the magnetic moment currents, the time derivative of the magnetic moment $d\mathbf{m}/dt$ and the hidden momentum b
oth depend crucially upon the velocity changes of the charged particles which produce the magnetic moment.  Thus when we calculated these quantities in the previous section, we made use of the explicit form for these magnetic-moment currents as arisi
ng from charges moving around a frictionless circular ring.

      In 1987 Solem[10] pointed out in the American Jounal of Physics that the response of a classical hydrogen atom to an external electric field was completely different from the polarization expected of an isotropic harmonic oscillator.  Whereas t
he familiar textbook model of atomic polarization (given by the harmonic oscillator) predicts a polarization in the direction of the external electric field, the response of a classical hyrogen is "strange;" the hydrogen atom begins by giving a polar
ization perpendicular to the direction of the applied electric field.  Furthermore, the external electric field leads to a change $d\mathbf{L}/dt$ in the angular momentum of the orbiting particle.  Solem's observation is the crucial beginning step in
 understanding the interaction of a point charge and a relativistic magnetic dipole moment.

\noindent{\bf c) The Changing Magnetic Moment}

     Our relativistic (to order $1/c^2$) model for a magnetic moment consists of noninteracting classical hydrogen atoms.  The external point charge $q$ is regarded as moving with velocity $\mathbf{v}_q$ past the magnetic moment.  We think of the hyd
rogen atoms as undergoing many orbital revolutions during the passage of the point charge and also regard each system as providing only a small perturbation on the other.  The passing charge is regarded as sufficiently far from the magnetic moment th
at the electrostatic field of the charge can be taken as uniform across the magnetic moment.  Thus the situation corresponds to that discussed in Solem's analysis.[10]

     The acceleration of a nonrelativistic electron in a classical hydrogen atom in a uniform external electric field $\mathbf{E}_0$ is given by Newton's second law with a force of Coulomb attraction to the nucleus and a force $e_b\mathbf{E}_0$ due t
o the external field
\begin{equation}
M_b\frac{d^2\mathbf{r}_b}{dt^2}=-\frac{e^2_b\mathbf{r}_b}{r^3_b}+e_b\mathbf{E}_0
\end{equation}
As noted above, the rate of change of the magnetic moment of the classical hydrogen atom is proportional to the rate of change of the angular momentum of the charge $e_b$ in its orbit around the heavy nucleus.  The rate of change of angular momentum
is proportional to the torque $\mathbf{\Gamma}_b$ applied to the circulating charge by the electric field, $\mathbf{\Gamma_b}=\mathbf{r}_b\times\mathbf{F}_b=\mathbf{r}_b\times e_b\mathbf{E}_0$.  Averaging over the orbital motion, we have
\begin{equation}
\left<\frac{d\mathbf{m}}{dt}\right>=\left<\frac{e_b}{2M_bc}\mathbf{\Gamma}_b\right>=\left<\frac{e^2_b}{2M_bc}\mathbf{r}_b\times\mathbf{E}_0\right>=\frac{e^2_b}{2M_bc}\left<\mathbf{r}_b\right>\times\mathbf{E}_0
\end{equation}

     Now the initial motion of the circulating charge $e_b$ is a circular orbit so that the average displacement of the charge from the nucleus vanishes $<\mathbf{r}_b>=0$.  In the rigid-circular-ring model of the previous section all the velocity ch
anges are symmetric on the opposite sides of the ring perpendicular to $\mathbf{E}_0$ and the quantity $<\mathbf{r}_b>\times\mathbf{E}_0$ remains zero despite the presence of the external electric field.  However, for the magnetic moment model of the
 classical hydrogen atoms, Solem's work tell us that this average displacement changes due to the applied electric field and $<\mathbf{r}_b>\times\mathbf{E}_0$ becomes non-zero.[27]

     For a charged particle $e$ in a Coulomb orbit, the displacement from the origin can be written in the form
\begin{equation}
\mathbf{r}=\frac{3}{2}\frac{\mathbf{K}}{(-2MH_0)^{1/2}}+\frac{1}{4H_0}\frac{d}{dt}\left(M(\mathbf{r}\times\mathbf{v})\times\mathbf{r}+M\mathbf{v}r^2\right)
\end{equation}
where $\mathbf{K}$ is the Laplace-Runge-Lenz vector[28]
\begin{equation}
\mathbf{K}=\frac{1}{(-2MH_0)^{1/2}}\left([\mathbf{r}\times(M\mathbf{v})]\times(M\mathbf{v})+Me^2\frac{\mathbf{r}}{r}\right)\end{equation}
and $H_0$ is the particle energy
\begin{equation}
H_0=\frac{1}{2}Mv^2-\frac{e^2}{r}
\end{equation}
Equation (38) can be verified by simply carrying out the time derivative appearing there, and then inserting the equation of motion $\mathbf{a}=-e^2\mathbf{r}/(Mr^3)$ for every appearance of the acceleration $\mathbf{a}=d^2\mathbf{r}/dt^2$.  The Lapl
ace-Runge-Lenz vector is constant in time for a Coulomb orbit, and the second term in Eq.(38) involving a time derivative shows how the displacement $\mathbf{r}$ varies in time; on averaging, the time-derivative term vanishes giving
\begin{equation}
<\mathbf{r}>=\frac{3}{2}\frac{\mathbf{K}}{(-2MH_0)^{1/2}}
\end{equation}

     In our model, the rate of change of the Laplace-Runge-Lenz vector due to the uniform field $\mathbf{E}$ is given by time-differentiation on the left- and right-hand sides of the definition Eq.(39) and the use of the equation of motion (36),
\begin{eqnarray*}
\frac{d\mathbf{K}}{dt}=M\left\{\left[\mathbf{r}\times\left(\frac{-e^2\mathbf{r}}{r^3}+e\mathbf{E}\right)\right]\times\mathbf{v}+(\mathbf{r}\times\mathbf{v})\times\left(\frac{-e^2\mathbf{r}}{r^3}+e\mathbf{E}\right)\right\}+Me^2\left[\frac{\mathbf{v}}{
r}-\frac{\mathbf{r}(\mathbf{r}\cdot\mathbf{v})}{r^3}   \right]
\end{eqnarray*}
\begin{equation}
=Me[-2\mathbf{r}(\mathbf{v}\cdot\mathbf{E})+\mathbf{E}(\mathbf{r}\cdot\mathbf{v})+\mathbf{v}(\mathbf{r}\cdot\mathbf{E})]
\end{equation}
This equation reminds us that the Laplace-Runge-Lenz vector is time-independent for a Coulomb orbit in the absence of the external electric field $\mathbf{E}_q=0$, but does indeed change due to an external electric field.  Now in our model for a rela
tivistic magnetic moment involving classical hydrogen atoms, we are interested in orbits which are slightly perturbed by the electrostatic field of the passing point charge.  Thus through first order in the perturbing field $\mathbf{E}$, we can obtai
n the time derivative of $\mathbf{K}$ in Eq.(42) by using the unperturbed steady-state orbit on the right-hand side.  For this steady-state motion, the average over the orbit gives $<x_iv_j+v_ix_j>=0$ or $<x_iv_j>=-<v_ix_j>=(1/2)<x_iv_j-v_ix_j>$ and
$<\mathbf{r}\cdot\mathbf{v}>=0$.  Thus the time-derivative of $\mathbf{K}$ in Eq.(42), when averaged over the orbital motion, can be rewritten using the same techniques for steady-state motions which we used earlier  to find
\begin{equation}
\frac{d\mathbf{K}}{dt}=\left<Me[-2\mathbf{r}(\mathbf{v}\cdot\mathbf{E})+\mathbf{E}(\mathbf{r}\cdot\mathbf{v})+\mathbf{v}(\mathbf{r}\cdot\mathbf{E})]\right>
=\frac{3}{2}Me[(\mathbf{r}\times\mathbf{v})\times\mathbf{E}]=3cM\mathbf{m}\times\mathbf{E}
\end{equation}

\noindent{\bf d) The New Electric Force Back on the Passing Charge}

     Now in our model for a relativistic magnetic moment involving classical hydrogen atoms, we are interested in orbits which begin as circular, but then are slightly perturbed by the electrostatic field of the passing point charge.  Initially, when
 the point charge is far away,  a circular orbit of the magnetic moment has $<\mathbf{r}_b>=0$, corresponding to a vanishing Laplace-Runge-Lenz vector $\mathbf{K}$.  However, the electrostatic field causes a change in $\mathbf{K}$, which gives a non-
zero value for $<\mathbf{r}_b>$, which then produces a torque on the orbit due to the electrostatic field of the passing charge, which then leads to a change of the magnetic moment, and so an electric field back at the position of the passing charge.
  Thus combining equations (4), (37), (41), (42), and noting that the electric field at the magnetic moment due to the charge $q$ is $\mathbf{E}_q(\mathbf{r}_m,t)_{\mathbf{r}_m=0}=q(-\mathbf{n}_q)/r^2_q$ where $\mathbf{r}_q=\mathbf{n}_qr_q$ is the di
splacement of the charge $q$ from the magnetic moment $\mathbf{m}$ at the origin, the electric force on the passing charge $q$ due to the changing magnetic moment is
\begin{eqnarray*}
<\mathbf{F}_{onq}(t)>=q<\mathbf{E}_m(\mathbf{r}_q,t)>=\frac{q}{c}\frac{d\mathbf{m}}{dt}\times\frac{(-\mathbf{n}_q)}{r^2_q}=\frac{1}{c}\frac{e_b}{2M_bc}\frac{d\mathbf{L}_m}{dt}\times\mathbf{E}_q
\end{eqnarray*}
\begin{eqnarray*}
=\frac{1}{c}\frac{e_b}{2M_bc}<\Gamma>\times\mathbf{E}_q=\frac{1}{c}\frac{e_b}{2M_bc}\left(<\mathbf{r}_m>\times e_b\mathbf{E}_q\right)\times\mathbf{E}_q
\end{eqnarray*}
\begin{eqnarray*}
=\frac{1}{c}\frac{e_b}{2M_bc}\left(\frac{3}{2}\frac{\mathbf{K}}{(-2MH_0)^{1/2}}
\times e_b\mathbf{E}_q\right)\times\mathbf{E}_q
\end{eqnarray*}
\begin{eqnarray*}
=\frac{1}{c}\frac{e_b}{2M_bc}\left(\frac{3}{2}\frac{1}{(-2MH_0)^{1/2}}\int_{-\infty}^tdt'\frac{d\mathbf{K}}{dt'}\times e_b\mathbf{E}_q\right)\times\mathbf{E}_q
\end{eqnarray*}
\begin{eqnarray*}
=\frac{1}{c}\frac{e_b}{2M_bc}\left(\frac{3}{2}\frac{1}{(-2MH_0)^{1/2}}\left[\int_{-\infty}^tdt'3cM_b\mathbf{m}\times\mathbf{E}_q(t')\right]\times e_b\mathbf{E}_q\right)\times\mathbf{E}_q
\end{eqnarray*}
\begin{equation}
=\frac{9e_b^2}{4c(-2M_bH_0)^{1/2}}\left(\left[\int_{\infty}^tdt'\mathbf{m}\times\mathbf{E}_q(t')\right]\times\mathbf{E}_q\right)\times\mathbf{E}_q
\end{equation}

     As a specific example of this effect, we consider a point charge $q$ moving with (unperturbed) motion along the line $x=d$, $z=0$, with $y=Vt$, past the hydrogen-atom magnetic moment located at the origin and oriented in the z-direction $\mathbf
{m}=\mathbf{k}m$.  Then according to Eq.(44), the (small) perturbing back electric force due to the changing magnetic moment is given by
\begin{eqnarray*}
<\mathbf{F}_{onq}(t)>=\frac{9e_b^2}{4c(-2M_bH_0)^{1/2}}\left(\left[\int_{\infty}^tdt'\mathbf{k}m\times\frac{-q(\mathbf{i}d+\mathbf{j}Vt')}{(d^2+V^2t'^2)^{3/2}}\right]\times\frac{-q(\mathbf{i}d+\mathbf{j}Vt)}{(d^2+V^2t^2)^{3/2}}  \right)\times\frac{-q
(\mathbf{i}d+\mathbf{j}Vt)}{(d^2+V^2t^2)^{3/2}}
\end{eqnarray*}
\begin{eqnarray*}
=\frac{-9e_b^2q^3m}{4cV(-2M_bH_0)^{1/2}}\left[\left\{\frac{\mathbf{j}}{d}\left(\frac{Vt}{(d^2+V^2t^2)^{1/2}}+1  \right)+\frac{\mathbf{i}}{(d^2+V^2t^2)^{1/2}}  \right\}\times\frac{\mathbf{i}d+\mathbf{j}Vt}{(d^2+V^2t^2)^{3/2}}\right]  \times\frac{\mathbf{i}d+\mathbf{j}Vt}{(d^2+V^2t^2)^{3/2}}
\end{eqnarray*}
\begin{equation}
=\frac{-9e_b^2q^3m}{4cV(-2M_bH_0)^{1/2}}\left[\frac{\mathbf{i}y-\mathbf{j}d}{(d^2+y^2)^3}\right]
\end{equation}
where in the last line we have used $y=Vt$.  We notice that the y-component of the force (45) reverses sign with the sign of the x-coordinate $d$, and hence is opposite for charges passing on opposite sides of the magnetic moment.  This suggests a cl
assical electromagnetic lag effect: particles which start out side-by-side will no longer be side-by-side after they pass on opposite sides of a magnetic moment.[29]

     We see that the electric force back on the passing point charge $q$ due to the magnetic moment is totally different from that suggested by the rigid-ring model.  Here in this relativistic model we find that the electric force on $q$ is i) odd in
 the charge $q$ and so reverses sign with $q$, ii)depends upon the magnetic moment $\mathbf{m}$ and reverses sign when $\mathbf{m}$ reverses sign, iii)is of order $1/c^2$ since in Gaussian units $\mathbf{m}$ is of order $1/c$, and iv) the y-component
 of the force changes directions for charges passing on opposite sides of the magnetic moment.

\noindent{\bf e) Motion of the Center of Energy of the Magnetic Moment}

     There have been several attempts to introduce equations of motion for magnetic moments which seem to depart from the familiar Newton's second law with the force given by the Lorentz force.  Indeed, there is considerable confusion over the term "
force" and the designation of just what is accelerating when dealing with relativistic systems.[30]  In the previous section we saw that our relativistic model for a magnetic moment gave a Lorentz force back on a passing charge which is completely di
fferent from that suggested by the traditional rigid-ring model.  In the present section we analyze the motion of the magnetic moment from a relativistic point of view.  Again we emphasize that the possiblity that the magnetic moment may change in ti
me is crucial in treating both the forces due to the magnetic moment and the motion of the magnetic moment itself.

     Our relativistic model for the magnetic moment is that of noninteracting hydrogen atoms which interact with the external charge through the Darwin Lagrangian[1,26]
\begin{equation}
L=\Sigma_a\left(\frac{1}{2}M_av_a^2+\frac{1}{8c^2}(v_a^2)^2\right)-\frac{1}{2}\Sigma_a\Sigma_{b\neq a}\frac{e_ae_b}{r_{ab}}+\frac{1}{2}\Sigma_a\Sigma_{b\neq a}\frac{e_ae_b}{2c^2r_{ab}}\left(\mathbf{v}_a\cdot\mathbf{v}_b+\frac{(\mathbf{v}_a\cdot\mathbf{r}_{ab})(\mathbf{v}_b\cdot\mathbf{r}_{ab})}{r_{ab}^2} \right)
\end{equation}
The first two terms involving the mass $M_a$ represent the expansion of the usual free-particle Lagrangian $L=-mc^2(1-v^2/c^2)^{1/2}$ while the terms involving the charges take the form of a spatial integral over $E^2-B^2$.  Specifically, the second
group of terms involving $e_ae_b$ is the electrostatic energy $1/(8\pi)\int E^2 dV$ and the last group of terms invoving $e_ae_bv_av_b$ corresponds to the magnetic energy $1/(8\pi)\int B^2 dV$, where only the finite cross terms in $\mathbf{E}_a\cdot\mathbf{E}_b$ and $\mathbf{B}_a\cdot\mathbf{B}_b$ are retained.[31]  The Euler-Lagrange equations of motion take the form,
\begin{equation}
0=\frac{d}{dt}\frac{\partial L}{\partial \mathbf{v}_a}-\frac{\partial L}{\partial \mathbf{r}_a}
\end{equation}
If we carry out the first time derivative in Eq.(46), then the equations of motion for each particle take the form of Newton's second law where the force is given by the Lorentz force,
\begin{equation}
\frac{d\mathbf{p}^{mechanical}_i}{dt}=e_i\mathbf{E}(\mathbf{r}_i,t) +e_i\left(\frac{\mathbf{v}_i}{c}\right)\times\mathbf{B}(\mathbf{r}_i,t)
\end{equation}
where $\mathbf{E}(\mathbf{r}_i,t)$ and $\mathbf{B}(\mathbf{r}_i,t)$ are the electric and magnetic fields due to the other charges through order $1/c^2$.    For a point charge $e$ at the origin, these fields are exactly as in Eq.(7) and (8),
\begin{equation}
\mathbf{E}=\frac{e\mathbf{n}}{r^2}\left(1+\frac{v^2}{2c^2}-\frac{3(\mathbf{v}\cdot\mathbf{n})^2}{2c^2}\right)-\frac{e}{2c^2r}[\mathbf{a}-(\mathbf{a}\cdot\mathbf{n})\mathbf{n}]
\end{equation}
\begin{equation}
\mathbf{B}=e\frac{\mathbf{v}}{c}\times\frac{\mathbf{n}}{r^2}
\end{equation}
where $\mathbf{r}=\mathbf{n}r$ is the displacement of the field point from the charge $e$, $\mathbf{v}$ is the charge velocity, and $\mathbf{a}$ the acceleration.  These field expressions from the Darwin Lagrangian correspond with those used by Page
and Adams[17] which follow from the usual expressions for the electromagnetic fields of a point charge[32] when expanded through order $1/c^2$.

     The Lorentz force on the magnetic moment is taken as the sum of the Lorentz forces on the individual charges $e_b$ and $e_b'$ of the magnetic moment.  We emphasize that when we refer to a "force" in this article, we are always referring to a Lor
entz force or the sum of Lorentz forces.  However, other authors use the term "force" to refer to other quantities.[1]

     Although we are thinking of a relativistic magnetic moment model which involves the superposition of many hydrogen atoms (and antiatoms), we will describe the interaction as though it involved exactly three particles.  The charge $e_b$ of mass $
M_b$ is in a Coulomb orbit around the heavy nucleus of charge $e_{b'}=-e_b$ and mass $M_{b'}$ while the charge $q$ passes at a large distance away.  Taking the origin at the center of energy of the unperturbed magnetic moment particles, we can discus
s the energy, linear momentum, and velocity of the center of energy for the magnetic moment system, the passing charge $q$, and the total system composed of both parts.  The energy $U_m/c^2$ of the magnetic moment is
\begin{equation}
\frac{U_m}{c^2}=M_b\left(1+\frac{v_b^2}{2c^2}\right)+M_{b'}\left(1+\frac{v_{b'}^2}{2c^2}\right)-\frac{e_b^2}{c^2r_{bb'}}
\end{equation}
which includes the particle rest energies, kinetic energies, and electrostatic energy. The linear momentum of the magnetic moment is
\begin{equation}
\mathbf{P}_m=M_b(1+\frac{v^2_b}{2c^2})\mathbf{v}_b+M_{b'}(1+\frac{v^2_{b'}}{2c^2})\mathbf{v}_{b'}+\frac{e_be_{b'}}{2c^2r_{bb'}}\{\mathbf{v}_b+\mathbf{v}_b'+[(\mathbf{v}_b+\mathbf{v}_{b'})\cdot\mathbf{n}_{bb'}]\mathbf{n}_{bb'}\}
\end{equation}
and so includes both the mechanical momenta of the particles $M_i(1+v^2_i/2c^2)\mathbf{v}_i$ and the electromagnetic field momentum $\frac{e_be_{b'}}{2c^2r_{bb'}}\{\mathbf{v}_b+\mathbf{v}_b'+[(\mathbf{v}_b +\mathbf{v}_{b'})\cdot\mathbf{n}_{bb'}]\mathbf{n}_{bb'}\}
$ involving both particles $e_b$ and $e_{b'}$. Finally, the position $\mathbf{X}_m$ of the center of energy of the magnetic moment through order $1/c^2$ is given by[1]
\begin{equation}
\frac{U}{c^2}\mathbf{X}_m=M_b\left(1+\frac{1}{2}\frac{v^2_b}{c^2}\right) \mathbf{r}_b+M_{b'}\left(1+\frac{1}{2}\frac{v^2_{b'}}{c^2}\right)\mathbf{r}_{b'}- \frac{e_be_{b'}}{c^2r_{bb'}}\frac{\mathbf{r}_b+\mathbf{r}_{b'}}{2}
\end{equation}
The first term on the right represents the rest energy  and kinetic energy of the mass $M_b$ located at the positon $\mathbf{r}_b$, the second is the contribution from the nucleus $b'$ and the last term on the right represents the electrostatic energ
y $e_be_{b'}/r_{bb'}$ between charges $e_b$ and $e_{b'}$ assigned to a location midway between the particles.
For the passing charge $q$, the energy divided by $c^2$ is
\begin{equation}
\frac{U_q}{c^2}=M_q\left(1+\frac{v_q^2}{2c^2}\right)
\end{equation}
the linear momentum is
\begin{equation}
\mathbf{p}_q=M_q\left(1+\frac{v_q^2}{2c^2}\right)\mathbf{v}_q
\end{equation}
and the position $\mathbf{r}_q$ of the center of energy is
\begin{equation}
\frac{U_q}{c^2}\mathbf{r}_q=M_q\left(1+\frac{v_q^2}{2c^2}\right)\mathbf{r}_q
\end{equation}
The total system energy divided by $c^2$ is the sum of the energies of the parts (magnetic moment $\mathbf{m}$ and point charge $q$) since on time-average the electrostatic energy between the neutral magnetic moment and the charge $q$ vanishes.  (The
 average electric quadrupole and higher terms will vanish when averaged together with the anti-atom contributions to the magnetic moment.)  Thus
\begin{eqnarray*}
\frac{U_{total}}{c^2}=\frac{U_m}{c^2}+\frac{U_q}{c^2}
\end{eqnarray*}
\begin{equation}
=M_b\left(1+\frac{v_b^2}{2c^2}\right)+M_{b'}\left(1+\frac{v_{b'}^2}{2c^2} \right)-\frac{e_b^2}{c^2r_{bb'}}+M_q\left(1+\frac{v_q^2}{2c^2}\right)
\end{equation}

     The total momentum includes the momentum contributions from the magnetic moment and the passing charge and also the linear momentum of the electromagnetic field arising from the magnetic fields of the magnetic moment charges and the electrostati
c field of the passing charge.  The electromagnetic field momentum involves terms of the form $e_bq\mathbf{v}_b/r_{bq}$ which do not vanish on averaging because the various orientations of $\mathbf{v}_b$ are associated with different distances $r_{bq
}$.  Accordingly the total linear momentum is
\begin{eqnarray*}
\mathbf{P}_{total}=\mathbf{P}_m+\mathbf{p}_q+\mathbf{P}_{em}
\end{eqnarray*}
\begin{eqnarray*}
=M_b(1+\frac{v^2_b}{2c^2})\mathbf{v}_b+M_{b'}(1+\frac{v^2_{b'}}{2c^2}) \mathbf{v}_{b'}+\frac{e_be_{b'}}{2c^2r_{bb'}}\{\mathbf{v}_b+\mathbf{v}_b' +[(\mathbf{v}_b+\mathbf{v}_{b'})\cdot\mathbf{n}_{bb'}]\mathbf{n}_{bb'}\}
\end{eqnarray*}
\begin{equation}
+M_q\left(1+\frac{v_q^2}{2c^2}\right)\mathbf{v}_q+\frac{e_bq}{2c^2r_{bq}} \left(\mathbf{v}_b+\frac{(\mathbf{v}_b\cdot\mathbf{r}_{bq})\mathbf{r}_{bq}}{r_{bq}^2}  \right)
\end{equation}
Since the time-average of the electrostatic potential energy between the passing charge $q$ and the neutral magnetic moment vanishes, the center of energy of the total system is given by the separate contributions from the magnetic moment and the cha
rge $q$,
\begin{eqnarray*}
\frac{U_{total}}{c^2}\mathbf{X}_{total}=\frac{U_m}{c^2}\mathbf{X}_m+ \frac{U_q}{c^2}\mathbf{r}_q
\end{eqnarray*}
\begin{equation}
=\left[M_b\left(1+\frac{1}{2}\frac{v^2_b}{c^2}\right)\mathbf{r}_b+M_{b'} \left(1+\frac{1}{2}\frac{v^2_{b'}}{c^2}\right)\mathbf{r}_{b'}- \frac{e_be_{b'}}{c^2r_{bb'}}\frac{\mathbf{r}_b+\mathbf{r}_{b'}}{2}\right]
+M_q\left(1+\frac{v^2_q}{2c^2}\right)\mathbf{r}_q
\end{equation}

     The net Lorentz forces between the magnetic moment and the passing charge $q$ are of order $1/c^2$, and hence any net transfer of energy must be of order $1/c^2$.  Therefore the energies divided by $c^2$, $U_m/c^2$ and $U_q/c^2$, can be regarded
 as constant when we negelct terms of order $1/c^4$.  Although the energies divided by $c^2$ may be regarded as constant, the locations of the energies do indeed change in order $1/c^2$, and the continuity of the changing energy locations is a signif
icant relativistic departure from nonrelativistic mechanics.  Thus whereas nonrelativistic mechanics has no local location for energy and allows energy to be freely transfered from one point to another distant point, special relativity insists on a l
ocal location and transfer of energy.  This idea is reflected in the constant motion of the center of energy for a closed relativistic system.  Thus taking the time-derivative of Eq.(59), we have
\begin{eqnarray*}
const=\frac{U_{total}}{c^2}\frac{d\mathbf{X}_{total}}{dt}=\frac{U_m}{c^2} \frac{d\mathbf{X}_m}{dt}+\frac{U_q}{c^2}\frac{d\mathbf{r}_q}{dt}
\end{eqnarray*}
\begin{eqnarray*}
=M_b(1+\frac{v^2_b}{2c^2})\mathbf{v}_b+M_{b'}(1+\frac{v^2_{b'}}{2c^2}) \mathbf{v}_{b'}+\frac{e_be_{b'}}{2c^2r_{bb'}}\{\mathbf{v}_b+\mathbf{v}_b' +[(\mathbf{v}_b+\mathbf{v}_{b'})\cdot\mathbf{n}_{bb'}]\mathbf{n}_{bb'}\}
\end{eqnarray*}
\begin{equation}
+M_b\mathbf{a}_b\cdot\mathbf{v}_b\frac{\mathbf{r}_b}{c^2}+M_{b'}\mathbf{a}_{b'}\cdot\mathbf{v}_{b'}\frac{\mathbf{r}_{b'}}{c^2}+M_q\left(1+\frac{v_q^2}{2c^2}\right)\mathbf{v}_q
\end{equation}
Now we recognize the first three terms following the last equal sign as the linear momentum of the magnetic moment and the last term as the linear momentum of the passing charge $q$.  However, there are two additional, unfamiliar terms involving the
accelerations $\mathbf{a}_b$, $\mathbf{a}_{b'}$ of the charges of the magnetic moment.  These terms can be rewritten using Newton's second law $\mathbf{F}=M\mathbf{a}$ in terms of the forces $\mathbf{F}_{onb}$, $\mathbf{F}_{onb'}$ acting on the charg
es $e_b$ and $e_{b'}$; since the terms are already of order $1/c^2$, we may use the nonrelativistic form of Newton's second law,
\begin{equation}
M_b\mathbf{a}_b\cdot\mathbf{v}_b\frac{\mathbf{r}_b}{c^2}=\left[\frac{d}{dt}\left(\frac{1}{2}M_bv_b^2\right)\right]\frac{\mathbf{r}_b}{c^2}=(\mathbf{F}_{onb}\cdot\mathbf{v}_b)\frac{\mathbf{r}_b}{c^2}
\end{equation}

  Then the constant motion of the total system center of energy in (60) takes the form
\begin{equation}
const=\frac{U_{total}}{c^2}\frac{d\mathbf{X}_{total}}{dt}=\frac{U_m}{c^2} \frac{d\mathbf{X}_m}{dt}+\frac{U_q}{c^2}\frac{d\mathbf{r}_q}{dt}=\left( \mathbf{P}_m+\mathbf{F}_{onb}\cdot\mathbf{v}_b\frac{\mathbf{r}_b}{c^2} +\mathbf{F}_{onb'}\cdot\mathbf{v}_{b
'}\frac{\mathbf{r}_{b'}}{c^2}\right)+\mathbf{p}_q
\end{equation}
The term $(\mathbf{F}_{onb}\cdot\mathbf{v}_b)\mathbf{r}_b/c^2$ represents the power introduced by the force $\mathbf{F}_{onb}$ on the charge $e_b$ moving with velocity $\mathbf{v}_b$ located at $\mathbf{r}_b$, and there is a similar term for $e_{b'}$
.  At some points in its orbit the charge $e_b$ receives energy from the field $\mathbf{E}_q$ and at other points it returns  this energy.  Thus the constant motion of the total system center of energy includes the motion of the center of energy of t
he magnetic moment system associated with the power introduced at the positions $\mathbf{r}_b$ and $\mathbf{r}_{b'}$ of the charges of the magnetic moment.  This effect is familiar from the student relativity problem involving the shift of the energy
 from one end of a plank to another by the energy transferred by a conveyor belt.[33]  This is the emphatic reminder that energy in relativity has a local existence and that accounting for the local transfer of energy is an integral part of the theor
y.

     Now in our simple system involving point charges described by the Darwin Lagrangian, one can show that the power flow within the magnetic moment given in Eq.(62) is equal to the linear momentum within the electromagnetic field.[1]  Thus using th
e nonrelativistic ($0$-order in $1/c$) equations of motion
\begin{equation}
M_i\mathbf{a}_i=\Sigma_{j\neq i}\frac{e_ie_j\mathbf{r}_{ij}}{r_{ij}^3}
\end{equation}
and then averaging over the unperturbed motion, the power-flow terms can be written as
\begin{eqnarray*}
\left<\mathbf{F}_{onb}\cdot\mathbf{v}_b\frac{\mathbf{r}_b}{c^2}+\mathbf{F}_{onb'}\cdot\mathbf{v}_{b'}\frac{\mathbf{r}_{b'}}{c^2}\right>=\left<\left( \frac{e_be_{b'}\mathbf{r}_{bb'}}{r_{bb'}^3}+\frac{e_bq\mathbf{r}_{bq}}{r_{bq}^3}\right)\cdot\mathbf{v}
_b\frac{\mathbf{r}_b}{c^2}+\left(\frac{e_{b'}e_b\mathbf{r}_{b'b}}{r_{b'b}^3}+\frac{e_{b'}q\mathbf{r}_{b'q}}{r_{b'q}^3}\right)\cdot\mathbf{v}_{b'}\frac{\mathbf{r}_{b'}}{c^2}\right>
\end{eqnarray*}
\begin{equation}
=-\frac{1}{c}\mathbf{m}\times\mathbf{E}_q(\mathbf{r}_m,t)=\frac{q}{c} \mathbf{A}_m(\mathbf{r}_q,t)=\mathbf{P}_{em}
\end{equation}
where we have used the same sort of averaging as described earlier in Section 4b.  We note that there is no average internal energy transfer within the atom itself between the orbiting charge $e_b$ and the heavy nucleus $e_{b'}$ in the unperturbed mo
tion; indeed, in our example where the unperturbed orbits are circular, the power transfer between the nucleus and the circulating charge vanishes at all times.[34]  Thus the energy transfer within the unperturbed magnetic moment is due entirely to t
he electrostatic field of the passing charge $q$.  Although the electrostatic field of $q$ delivers no average energy to the magnetic moment, this electrostatic field does bring about a redistribution of energy within the magnetic moment.  In the rel
ativistic Darwin Lagrangian where the nonrelativistic forces are Coulomb forces, the contribution to the uniform motion of the center of energy of the magnetic moment is equal to the linear momentum in the electromagnetic field.  Thus using equation
(64), the uniform motion of the center of energy for the total system in Eq.(62) can be written in the form[1]
\begin{equation}
const=\frac{U_{total}}{c^2}\frac{d\mathbf{X}_{total}}{dt}=\frac{U_m}{c^2} \frac{d\mathbf{X}_m}{dt}+\frac{U_q}{c^2}\frac{d\mathbf{r}_q}{dt}=(\mathbf{P}_m+\mathbf{P}_{em})+\mathbf{p}_q
\end{equation}
which takes the same form as the statement of conservation of linear momentum for the system but where actually both the momentum of the magnetic moment and the electromagnetic field momentum terms are associated with the motion of the center of ener
gy of the magnetic moment.

     We emphasize that the internal transfer of energy between different locations of the magnetic moment, which gives a motion of the center of energy of the magnetic moment, represents an instabilty of the magnetic moment system.  The energy distri
bution within the magnetic moment is being changed, as seen in the orbit distortion noted by Solem for the classical hydrogen atom.[35]  A stable system will react in such a way as to prevent this internal redistribution of energy.  In our relativist
ic example here, the orbit of the magnetic moment charge $e_b$ is altered in just such a way as to give a changing magnetic moment which puts an electric force back on the passing charge.  In addition, the changing magnetic moment acts to decreases t
he change in linear momentum in the electromagnetic field and also to decrease the change of the velocity of the center of energy of the magnetic moment due to the internal energy transfer.

     If we differentiate equation (65) with respect to time, then we obtain the familiar conservation of linear momentum for the system of the magnetic moment and external charge as well as a statement about the accelerations of the centers of energy
 of the charge $q$ and the magnetic moment $\mathbf{m}$,
\begin{equation}
0=\frac{U_{total}}{c^2}\frac{d^2\mathbf{X}_{total}}{dt^2}= \frac{U_m}{c^2}\frac{d^2\mathbf{X}_m}{dt^2}+\frac{U_q}{c^2}\frac{d^2 \mathbf{r}_q}{dt^2}=\frac{d}{dt}\left(\mathbf{P}_m+\mathbf{p}_q+\mathbf{P}_{em}  \right)
\end{equation}
Since for the point charge $q$ we have $(U_q/c^2)(d^2\mathbf{r}_q/dt^2)=d\mathbf{p}_q/dt$, it follows that
\begin{equation}
\frac{U_m}{c^2}\frac{d^2\mathbf{X}_m}{dt^2}=\frac{d}{dt}\left(\mathbf{P}_m+\mathbf{P}_{em}  \right)
\end{equation}

     Now in equation (12) of Section 3a, we saw that the sum of the Lorentz forces on the charge and the magnetic moment satisfied, $\mathbf{F}_{onm}+\mathbf{F}_{onq}=-d\mathbf{P}_{em}/dt$ while $\mathbf{F}_{onq}=d\mathbf{p}_q/dt$ so that
\begin{equation}
\frac{U_m}{c^2}\frac{d^2\mathbf{X}_m}{dt^2}=\frac{d}{dt}\left(\mathbf{P}_m+\mathbf{P}_{em}\right)=\mathbf{F}_{onm}+\frac{d\mathbf{P}_{em}}{dt}=\mathbf{F}_{onm}+\frac{\partial}{\partial t}\left(\frac{\mathbf{E}_q\times\mathbf{m}}{c}  \right)
\end{equation}
This is just the equation of motion for a current-loop magnetic moment given by Haus and Penfield[36].

\noindent{\bf 6. PROPOSED INTERACTION OF A POINT CHARGE AND A MAGNET}

\noindent{\bf a) Discussion of the Equations of Motion for a Magnetic Moment}

In our relativistic model of a magnetic moment using the Darwin Lagrangian, the mass times the acceleration of the center of energy of the magnetic moment moves as given in Eq.(68).  This equation corresponds to exactly the expression given by Haus a
nd Pennfield[36].  However, the equation does not actually give the acceleration of the magnetic moment center of energy because it does not determine $d\mathbf{m}/dt$, the time-rate-of-change of the magnetic moment.  Only a detailed model for the ma
gnetic moment will provide this information, just as illustrated for our hydrogen-atom magnetic moment model.

    It appears that many researchers are unaware of the possible significant alteration in the magnetic moment, and therefore rewrite the equation as though the magnetic moment were constant.  In this case the equation takes the form
\begin{equation}
\frac{U_m}{c^2}\frac{d^2\mathbf{X}_m}{dt^2}=\mathbf{F}_m-\frac{\mathbf{m}}{c}\times\frac{\partial\mathbf{E}_q}{\partial t}
\end{equation}
which is the form given by Aharonov, Peale, and Vaidman.[5]  This is also the equation given for the force on a magnetic moment formed from two magnetic monopoles,[38] where we would not expect any variation of the magnetic moment due to interaction
with a passing charge.  However, this formualation is not valid for a relativistic current-loop magnetic moment where $\mathbf{m}$ will change with time.

  The electromagnetic field momentum $\mathbf{P}_{em}=(q/c)\mathbf{A}_m(\mathbf{r}_q,t)=-(\mathbf{m}(t)/c)\times\mathbf{E}_q(\mathbf{r}_m,t)$ changes for two reasons: because of the relative motion of $q$ and $\mathbf{m}$ and because of the change of
 the magnetic moment $\mathbf{m}$.  In our relativisitic hydrogen-atom magnetic moment model, the magnetic moment $\mathbf{m}$ changes in such a way as to decrease the change in electromagnetic field momentum and to provide an unexpected force on the
 passing charge.  We propose that a stable multiparticle magnetic moment will act not so as to counteract this behavior but rather such as to complete it, so as to make make negligible the internal transfer of energy from one part of the magnetic mom
ent to another while providing a force back on the passing charge which is the Newton's third law reaction to the magnetic Lorentz force of the passing charge on the magnetic moment.  Thus what we are proposing is that the magnetic moment changes in
just such a fashion that
\begin{equation}
\frac{d\mathbf{P}_{em}}{dt}=\frac{d}{dt}\left(\frac{q}{c}\mathbf{A}_m( \mathbf{r}_q,t)\right)=-\frac{d}{dt}\left(\frac{\mathbf{m}}{c}\times\mathbf{E}_q(\mathbf{r}_m,t) \right)=0
\end{equation}
This corresponds to no change in the electromagntic field linear momentum, and also, because this field momentum is related to the internal shift in the energy of the magnetic moment, to no shift in the internal energy distribution of the magnetic mo
ment.  But then the equation of motion of the magnetic moment becomes the old familiar
\begin{equation}
\frac{U_m}{c^2}\frac{d^2\mathbf{X}_m}{dt^2}=\mathbf{F}_m
\end{equation}
where the acceleration of the center of energy of the magnetic moment is connected to the Lorentz force on the magnetic moment by Newton's second law $\mathbf{F}=M\mathbf{a}$.  Furthermore, Newton's third law is valid and equations (66) and (68) can
be rewritten as
\begin{equation}
\frac{d\mathbf{P}_m}{dt}=\mathbf{F}_{onm}=M_m\frac{d^2\mathbf{X}_m}{dt^2}=-\frac{U_q}{c^2}\frac{d^2\mathbf{r}_q}{dt^2}=-\mathbf{F}_{onq}=-\frac{d\mathbf{p}_q}{dt}
\end{equation}
Thus the change of a multiparticle magnetic moment in this fashion brings us back to familiar basic physics, but with also the unfamiliar idea that the Lorentz force on a passing charge $q$ includes the electric field arising from the changing magnet
ic moment.

\noindent{\bf b) Analogy with a Conductor in Electrostatics}

     There is a close analogy between the situation for a point charge outside a conductor and a point charge outside a neutral long solenoid.  Both involved the interaction of a point charge with a stable, multiparticle structure.  Just as the unper
turbed conductor has no electric or magnetic field outside, so the long solenoid has no electric or magnetic fields outside.  Therefore a point charge outside an unperturbed conductor experiences no force, just as a point charge outside an unperturbe
d solenoid experiences no force.  Also, the unperturbed conductor experiences no net force due to the external charge, just as the unperturbed solenoid experiences no net energy transfer (0-order in $1/c$) due to the electrostatic field of the passin
g charge.  However, the electrostatic field of the point charge puts different forces on the different charges of the conductor, leading to a transfer of momentum from one part (positive charges) of the conductor to another (negative charges).  This
leads to an instability which is eventually stabilized by the conductor so as to stop the net transfer of momentum from one part of the conductor to another while producing net forces between the conductor and the external charge which satisfy Newton
's third law.  In the case of the solenoid, the external electrostatic field puts forces on the currents which lead to a transfer of energy from one part (where currents approach the passing charge $q$) of the solenoid to another (where currents move
 away from $q$).  In a relativistic system, energy has a localized existence.  The transfer of energy from one part of the solenoid to another represents a shift of the center of energy corresponding to an instability.  We suggest that this instabili
ty is eventually stabilized by the solenoid so as to halt the transfer of energy from one part of the solenoid to another while producing net forces between the solenoid and the external charge which satisfy Newton's third law.  For the case of a con
ductor, the beginnings of such stabilizing behavior can be seen for a point charge near an isotropic dipole oscillator, which oscillator could be used as a polarizable building element for a conducting wall.  In this article, we have shown the beginn
ings of such stabilizing behavior for a hydrogen-atom model for a magnetic moment, which hydrogen-atom can be used as a building element for a solenoid.

\noindent{\bf 7. THE AHARONOV-BOHM AND AHARONOV-CASHER PHASE SHIFTS}

\noindent{\bf a) Connection with Charge-Magnet Interaction}

     The Aharonov-Bohm phase shift involves the quantum interference pattern shift occurring when electrons (point charges) pass on opposite sides of a solenoid or toroid (stacks of magnetic moments) and then are recombined to give a particle interfe
rence pattern.[6,7]  The Aharonov-Casher phase shift involves neutrons (magnetic moments) passing on opposite sides of a line charge (stack of point charges).[8,9]  Both interference pattern shifts are presented in the theoretical literature and in q
uantum mechanics texts books as examples of quantum topological effects occurring in the absence of classical electromagnetic forces.  Actually however, the classical electromagnetic forces between the charges and the magnetic moments have probably b
een completely misunderstood because of the nonrelativistic classical electromagnetic models for magnetic moments which appear in the instructional and research literature.  Special relativity has many paradoxes for the unwary, and this is probably o
ne of them.[39]  When a relativistic model for a magnetic moment is used, then the forces between a point charge and a magnetic moment are completely different from those of the usual nonrelativistic model.  Indeed, the relativistic force calculation
s lead to the likelihood that the observed phase shifts are based upon classical electromagnetic forces[40] and are not due to any new quantum topological effect which represents a further departure from classical physics.

     It has been pointed out in earlier work[40] that the classical electromagnetic analysis for both the Aharonov-Bohm and Aharonov-Casher phase shifts reduces to an understanding of the interaction of a point charge and magnetic dipole moment.  Thu
s by stacking magnetic moments appropriately one can pass from the basic charge-moment interaction over to the charge-solenoid or charge-toroid interactions which are involved in the Aharonov-Bohm phase shift.  By stacking point charges appropriately
, one can pass from the basic charge-moment interaction over to the line charge-moment interaction of the Aharonov-Casher phase shift.  However, the current understanding of the classical electromagnetic forces between a point charge and a solenoid i
s such that classical forces have been rejected as the basis for the Aharonov-Bohm and Aharonov-Casher phase shifts.  The current view in physics turns for understanding not to relativistic classical electromagnetic inteactions but rather to an ad ho
c, non-local quantum topological effect suggested by Aharonov and Bohm at the same time that they proposed the experiment.  I believe that topological view is singularly unhelpful and may not even give all the experimentally detectable aspects of the
 effect.[41]  The quantum calculation for the Aharonov-Bohm phase shift makes use of the nonrelativistic Schroedinger equation.  However, the effect is actually a relativistic shift of order $1/c^2$, and, since the corresponding classical problem has
 never been given an adequate Hamilton-Jacobi formulation, there remains considerable uncertainty as to what classical situation is actually represented by the Schroedinger equation for a charged particle passing a long solenoid.

\noindent{\bf b) The Claimed Quantum Topological Effect vs Classical Forces}

     For nearly thirty years I have suggested that the field of the passing charge $q$ must cause changes in the currents of the magnetic moment (or solenoid or toroid) and that the changing currents must produce electric fields back at the passing c
harge which might account for the experimentally observed Aharonov-Bohm phase shift.[42]  However, these suggestions have been almost universally ignored, and my recent  manuscripts have been rejected by The Physical Review.[43]  The referee(s) at Th
e Physical Review claim that my views are impossible because of the following argument.[44]  There are no electric or magnetic fields outside the winding of a long unperturbed solenoid.  Therefore the fields back at the passing charge must be due to
a perturbation which will be of order $qe_b$ where $e_b$ is the charge of a particle in the solenod.  These perturbations then produce an electric field of order $qe_b^2$ which then produces a force of order $q^2e_b^2$ back at the passing charge.  In
deed, in the present article, we saw that the rigid-ring model for a magnetic moment gave a Faraday induced electric field and force back at the passing charge $q$ of order $q^2e_b^2/c^4$ which was independent of the magnitude of the magnetic moment.
  This is totally different from forces necessary to account for the Aharonov-Bohm phase shift where the back force must depend linearly on $q/c$ and on the magnetic moment $\mathbf{m}$.

     The question of the interaction of a magnetic moment with an external charge came up even more emphatically in connection with the Aharonov-Casher phase shift which these authors claimed was the "dual" of the Aharonov-Bohm phase shift.  In this
case the authors claimed that there was no force on a magnetic moment passing a line charge $\lambda$.[8]  Actually there is indeed a net Lorentz force on a magnetic moment modeled as a current loop[45]; however, Aharonov and Casher used a magnetic m
oment model involving fixed magnetic charges, and for that model there is no force for their situation.  It was shown that when the magnetic dipole moment was modeled as a current loop and Newton's second law $\mathbf{F}=M\mathbf{a}$ together with th
e Lorentz force on a magnetic moment $\mathbf{F}_m=\nabla(\mathbf{m}\cdot\mathbf{B}_{\lambda})$ was used, then Aharonov-Casher phase shift could be understood as based upon a classical electromagnetic lag effect arising from classical electromagnetic
 forces.[45]  In the ensuing controversy  Aharonov, Pearle, and Vaidman conceded that a magnetic moment, modeled as a current loop, indeed experiences the suggested net Lorentz force.[46]  Nevertheless, because of "hidden momentum in magnets," they i
nsisted that the center of energy of the magnetic moment still moves in space as though it experiences no Lorentz force.  In other words, the electric field of the line charge $\lambda$ causes confusion behind the scenes in the passing magnetic momen
t, but the center of energy of the magnetic moment and of the line charge are both unaffected.  General experience suggests that nature does not behave this way.  Aharonov, Pearl, and Vaidman introduced the equation of motion (69) to justify their as
sertion of uniform motion, and their view seems to have prevailed in the physics community.[47]

     In the present article, we readress this long-standing controvery at its foundations. For a general magnet we are unable to carry through the full calculation of $d\mathbf{m}/dt$.  However, in the case of our relativistic magnetic moment model i
nvolving classical hydrodgen atoms, we showed that a surprising, unanticipated, classical electromagnetic force indeed occurs.

     In the traditional, rigid-loop model for a magnetic moment (which is not a relativistic model), we saw that Faraday induction produced a back force at the passing charge $\mathbf{F}_{on q}=(q/c)(\mathbf{n}_q/r^2_q)\times(d\mathbf{m}/dt)$ proport
ional to $q^2e^2/c^4$ where $d\mathbf{m}/dt$ arose from the changing magnetic flux due to the magnetic field $\mathbf{B}_q(0)=(q/c)[\mathbf{v}_q\times(-\mathbf{n}_q)]/r^2_q$ and two more factors of $1/c$ are added by Faraday's law $Emf=-(1/c)(d\phi/d
t)$ (where $\phi$ is the magnetic flux), and the definition of a magnetic dipole moment.  However, in the relativistic (to order $v^2/c^2$) model of a magnetic moment, there is a totally different mechanism which gives rise to a changing magnetic mom
ent.  This is the changing magnetic moment which arises from Solem's fascinating observation regarding the strange polarization of the classical hydrogen atom.  A nonrelativistic classical hydrogen atom does not become polarized like a harmonic oscil
lator.  On the contrary, the nonrelativistic atom changes its orbit and its angular momentum due to a weak external electric field.  Thus the magnetic moment $\mathbf{m}=[e/(2Mc)]\mathbf{L}$ of the nonrelativistic orbit changes.  The magnetic moment
changes not due to Faraday induction from the magnetic field of the passing charge $q$ but rather from the nonrelativistic alteration of the orbit due to the electric field of the passing charge $q$.  The electric field back at the passing charge is
not the nonrelativistic result of the rigid loop model, but rather is or order $q^3e^3\omega/c^2$ where $\omega$ is the angular velocity of the charge $e$ in Coulomb orbit which produces the magnetic moment.  Thus this electric force on the passing c
harge indeed depends on the magnetic moment $eR^2\omega/(2c)$, which is in total contrast with the Faraday-induced force for a rigid-loop model of a magnetic moment.  In this relativistic case, the back force is odd in the charge $q$ of the passing p
article.  In the relativistic case there is indeed a relative classical lag effect which reverses for charges passing on opposite sides of the magnetic moment.  There is indeed the qualitative basis for understanding the Aharonov-Bohm and Aharonov-Ca
sher phase shifts as based upon classical electromagnetic forces.

     In our simple relativistic model, we assumed that the classical atoms forming the magnetic moment did not interact.  If they did interact, then the changing magnetic moment $d\mathbf{m}/dt$ would be different.  What has been suggested is that fo
r a realistic multiparticle magnetic moment $d\mathbf{m}/dt$ assumes exactly the value such the magnetic moment moves under just the net Lorentz force produced by the passing charge.  If this is indeed the case, then the Aharonov-Bohm and Aharonov-Ca
sher phase shifts would be understood as arising from a classical electromagnetic lag effect arising from classical electromagnetic forces between the point charge and the magnetic moment.  In the case of the Aharonov-Bohm effect, the passing charge
is not moving in a field-free region but rather experiences a net Lorentz force $\mathbf{F}_q=[(q\mathbf{v}_q/c)\cdot\nabla_q]\mathbf{A}_m(\mathbf{r}_q)$ which is equal in magnitude and opposite in direction to the Lorentz force on the solenoid.[40]
 The Lorentz force on the solenoid is due entirely to the familiar magnetic force $\mathbf{F}_m=\Sigma _b(e_b\mathbf{v}_b/c)\times\mathbf{B}_q(\mathbf{r}_b,t)$ arising from the magnetic field of the  passing charge, while the Lorentz force on the pas
sing charge in the case of an infinite solenoid or toroid is due entirely to the changes $d\mathbf{m}/dt$ in the magnetic moments of which the solenoid can be regarded as composed.[48]

\noindent{\bf c) The Connections Between Classical and Quantum Physics}

     During the second quarter of the twentieth century there developed a rough sense of the connections between classical and quantum physics which eventually gave expression to the stochastic interpretations of quantum mechanics[49] and to stochast
ic electrodynamics.[50]  This rough understanding is challenged by the Aharonov-Bohm effect which purports to be a quantum topological effect occurring in the absence of classical electromagnetic forces.  This new point of view has entered the textbo
ok literature.  Griffiths, with his usual clarity, gives one of the sharpest expressions of this new point of view.  In his quantum text he writes, "What are we to make of the Aharonov-Bohm effect? Evidently our classical preconceptions are simply \t
extit{mistaken}: There \textit{can} be electromagnetic effects in regions where the fields are zero."[51]  I believe that this point of view is probably in error.

     For thirty years there have been attempts to understand the experimentally-observed Aharonov-Bohm and Aharonov-Casher phase shifts within the older semiclassical understanding of the connection between classical and quantum theories.[40,41,42]
The classical electromagnetic phenomena associated with these phase shifts have provided fertile clues to aspects of classical electromagnetic theory which had been previously overlooked.  And all of the new results obtained tend to increase the like
lihood of a semiclassical explanation of these phase shifts.  Thus persistence of the phase shift when the solenoid was surrounded by a conductor[52] led to the realization that electromagnetic velocity fields have an algebraic fall off in conductors
 which is quite different from the exponential damping of electromagnetic wave fields, and also to the result that the magnetic velocity fields penetrate conductors independently of the conductivity at low velocities.[53]  These results have been acc
epted into the textbook literature.[54]  The decrease of the magnetic velocity fields at high velocities[55] led to the discovery of a new time-integral invariant of the magnetic velocity fields of a passing charge which is entirely independent of an
y ohmic conductors which may be present,[56] and only this invariant time-integral enters the classical basis for the Aharonov-Bohm phase shift.[40]  The frequent objections that the phase shift could not be caused by classical electromagnetic forces
  has led to the rethinking of the interaction of a point charge and a magnetic moment presented in the present article.  Thus attempts to give a semiclassical understanding of the Aharonov-Bohm and Aharonov-Casher phase shifts have led to a new unde
rstanding of the penetration of electric and magnetic velocity fields into ohmic conductors and to a new understanding of the point charge-magnetic moment interaction.

\noindent{\bf 8. CLOSING SUMMARY}

     In this article we consider two different models for magnetic moments.  First we discuss a commonly used magnetic moment model which is not relativistic.  This model involves a single charged particle $e_b$ sliding on a frictionless rigid circul
ar ring of opposite charge.  The rigid structure immediately betrays the nonrelativistic character of the model.  We find that a point charge q passing this rigid-ring model produces a Faraday induction force back on the passing charge which is of or
der $(e_bq)^2/c^4$.  This force is negligible since we retain forces only through order $1/c^2$.  This inadequate model, which is not relativisitic, has no possibility of giving accurate forces at the relativistic level between the passing charge and
 the magnetic dipole moment.  The model does not give forces remotely appropriate to provide a classical explanation of the Aharonov-Bohm phase shift.

     The second model satisfies the requirements of special relativity through order $1/c^2$, but is not realistic for a many-particle magnet.  This magnetic-moment model consists of many classical hydrogen atoms (and anti-atoms) superimposed, the at
oms interacting with an exteral charge q but not with each other.  When we apply Solem's analysis for the strange polarization of the classical hydrogen atom, we find totally new aspects for the interaction of a point charge and a magnetic moment.  T
here is a change in the magnetic moment of the hydrogen atom which occurs due to the nonrelativistic interaction of the hydrogen atom with an external electric field.  We then follow the analysis of Coleman and Van Vleck using the Darwin Lagrangian a
nd note the role of the hidden momentum.  It seems interesting that Coleman and Van Vleck are aware of the role  of the electric field of the passing charge $q$ in influencing the hidden momentum of the magnet in order $1/c^2$, but do not seem aware
of the role of the electric field in causing a change in the magnetic moment; Solem is aware of the influence of the electric field in changing the angular momentum of the nonrelativistic hydrogen atom but does not comment on the associated change in
 the magnetic moment.  The changing magnetic moment leads to electric forces back on a passing charge which are of order $q^3e_b^3\omega/c^2$.  These forces have the qualitative behavior to fit with a classical-based explanation of the Aharonov-Bohm
phase shift.  i) The forces are of order $1/c^2$, not $1/c^4$.  ii) The forces involve odd powers of the passing charge.  iii) The forces are linear in the magnetic moment.  iv) The forces in the direction of motion are reversed on opposite sides of
the solenoid.

     The third part of this article considers what behavior we might expect from a more realistic magnetic moment (such as a solenoid or bar magnet) where all the charged particles interact.  In this case we argue from the analogy of the behavior of
a conductor and the relativistic magnetic moment model above.  We suggest that a realistic magnet acts so as to stop the redistribution of the mass-moments of the system.  This criterion is equivalent to assuming that the rate of change of the magnet
ic moment is precisely such as to give vanishing change in the electromagnetic field momentum between the magnet and the passing charge.  It also is precisely such as to give equal and opposite Lorentz forces on the magnet and the external charge, an
d it also is precisely such as to have the center of energy of the magnet and of the passing charge move according to Newton's second law $\mathbf{F}=M\mathbf{a}$ where $\mathbf{F}$ is the Lorentz force on each object.  The criterion is also precisel
y such as to give a classical electromagnetic lag effect as the basis for the Aharonov-Bohm and Aharonov-Casher phase shifts while restoring the usual connection between classical and quantum theories for these phenomena.

\noindent{\bf Acknowlegement}

I would like to thank the Drs. L. James Swank and M. Jean Swank for a useful discussion.  The analysis given in the present article is deeply indebted to the published work of J. C. Solem [Am. J. Phys. {\bf 55}, 906 (1987)] of L. C. Biedenharn, L. S.
 Brown, and J. C. Solem [Am. J. Phys. {\bf 56}, 661 (1988)], and of S. Coleman and J. H. Van Vleck [Phys. Rev. {\bf 171}, 1370 (1968)].


\begin{references}
\bibitem{} S. Coleman and J.H.Van Vleck, "Origin of 'Hidden Momentum Forces' on Magnets," Phys. Rev. {\bf 171}, 1370-1375 (1968).

\bibitem{4} W. Shockley and R.P. James, "Try Simplest Cases' Discovery of 'Hidden Momentum' forces on 'Magnetic Currents'," Phys. Rev. Letters {\bf 18}, 876-879 (1967).

\bibitem{ } O. Costa de Beauregard, "A New Law in Electrodynamics," Phys. Letters {\bf 24A}, 177-178 (1967).

\bibitem{ } A recent sampling includes, D. J. Griffiths, "Dipoles at rest," Am. J. Phys. {\bf 60}, 979-987 (1992); L. Vaidman, "Torque and force on a magnetic dipole," Am. J. Phys. {\bf 58}, 978-983 (1990); V. Namias, "Electrodynamics of moving dipol
es: The case of the missing torque," Am. J. Phys. {\bf 57}, 171-177 (1989); T.H. Boyer, "Force on a magnetic dipole," Am. J. Phys. {\bf 56}, 688-692 (1988).

\bibitem{ } See, for example, Y. Aharonov, P. Pearle, and L. Vaidman, "Comment on 'Proposed Aharonov-Casher effect: Another example of an Aharonov-Bohm effect arising from a classical lag," Phys. Rev. A {37}, 4052-4055 (1988); L. Vaidman, Ref. 4.

\bibitem{1} Y. Aharonov and D. Bohm, "Significance of Electromagnetic Potentials in Quantum Theory," Phys. Rev. {\bf 115}, 485-491 (1959).

\bibitem{2} R.G. Chambers, "Shift of an Electron Interference Pattern by Enclosed Magnetic Flux," Phys. Rev. Lett. {\bf 5}, 3-5 (1960) or G. Mollenstedt and W. Bayh, "Messung der kontinuierlichen Phasenschiebung von Elektronenwellen in kraftfeldfreie
n Raum durch das magnetishce Vektorpotential einer Luftspule," Naturwissenschaften {\bf 49}, 81-82 (1962).  See also the reviews by S. Olariu and I.I. Popescu, "The quantum effects of electromagnetic fluxes," Rev. Mod. Phys. {\bf 57}, 339-436 (1985),
 and by M. Peshkin and A. Tonomura, \textit{The Aharonov-Bohm Effect (Lecture notes in Physics 340)} (Springer-Verlag, New York 1989).

\bibitem{ } Y. Aharonov and A. Casher, "Topological Quantum Effects for Neutral Particles," Phys. Rev. Lett. {\bf 53}, 319-321 (1984).

\bibitem{7} A.W. Overhauser and R. Colella, Phys. Rev. Letters {\bf 33}, 1237 (1974).  R. Colella, A.W. Overhauser, and S.A. Werner, Phys. Rev. Lett. {\bf 34}, 1472 (1974).

\bibitem{ } J. C. Solem, "The strange polarization of the classical atom,"  Am. J. Phys. {\bf 55}, 906-909 (1987); L. C. Biedenharn, L. S. Brown, and J. C. Solem, "Comment on 'The strange polarization of the classical atom,'"  Am. J. Phys. {\bf 56},
661-663 (1988).  See also, J. C. Solem, Found. Phys. {\bf 27}, 1291-1306 (1997), "Variations on the Kepler Problem."

\bibitem{11} See, for example, J.D. Jackson, \textit{Classical Electrodynamics, 2nd. ed.} (Wiley, New York 1975), p. 172.

\bibitem{ } See, for example, J.D. Jackson, Ref.11,   p. 185.

\bibitem{ } See, for example, J.D. Jackson, Ref. 11,  p. 216.

\bibitem{ } See, for example, J.D. Jackson, Ref. 11,   p. 182.  The radiation fields are not included in this quasistatic treatment.

\bibitem{ } See, for example, J.D. Jackson, Ref. 11,  p. 594.

\bibitem{ } See, for example, J.D. Jackson, Ref. 11,  p. 185.

\bibitem{ } See, for example, L. Page and N. I. Adams, "Action and Reaction Between Moving Charges," Am. J. Phys. {\bf 13}, 141-147 (1945).

\bibitem{ } S. Coleman and J. H. Van Vleck, Ref.1,  p. 1373, Eq.(15).

\bibitem{ } Y. Aharonov, P. Pearle, and L. Vaidman, Ref. 5.

\bibitem{ } B. D. H. Tellegen, "Magnetic-Dipole Models," Am. J. Phys. {\bf 30}, 650-652 (1962); H.A. Haus and P. Penfield Jr, "Force on a Current Loop," Phys. Letters {\bf 26A}, 412-413 (1968).

\bibitem{ } See, for example, J.D. Jackson, Ref. 11, p. 594.

\bibitem{ } See, for example, L. Page and N. I. Adams, Ref. 17.

\bibitem{ }  These are precisely the results which are often quoted in claims that the Aharonov-Bohm phase shift can not possibly arise as the result of classical electromagnetic forces.

\bibitem{ } D. J. Griffiths, \textit{Introduction to Electrodynamics} (Prentic Hall, Upper Saddle River, NJ, 1999), 3rd ed., pp. 520-521.

\bibitem{ } This linear momentum must be associated with the rigid-wire constraint.  Clearly it is not provided by the electrostatic field $\mathbf{E}_q(0,t)$.  See T.H. Boyer, "Classical Electromagnetism and the Aharonov-Bohm Phase Shift," Found. Ph
s. {\bf 30}, 907-932 (2000), pp. 926-928.

\bibitem{ } See, for example, J.D. Jackson, Ref. 11, p. 595.

\bibitem{ }  Our analysis, although given in a different form, depends heavily on the insights provided by Solem, and by Biedenharn, Brown, and Solem in Ref. 10.

\bibitem{ } H. Goldstein, \textit{Classical Mechanics} (Addison-Wesley, Reading, MA, 1981), 2nd ed., pp. 102-104. The sign and normalization have been chosen as in Ref. 10.

\bibitem{ } See, for example, T.H. Boyer, "Classical Electromagnetic Deflections and Lag Effects Associated with Quantum Interference Pattern Shifts: Considerations Related to the Aharonov-Bohm Effect," Phys. Rev. D {\bf 8}, 1679-1693 (1973).

\bibitem{ } S. Coleman and J. H. Van Vleck mention various definitions in their article, Ref. 1, pp. 1373-1374.

\bibitem{ } The integrals over the electromagnetic fields give the quantities which appear in the Darwin Lagrangian.  See, for example, L. Page and N. I. Adams, Ref. 17.

\bibitem{ } The exact electromagnetic fields due to a point charge are given in J. D. Jackson's text (Ref. 11) on p. 657.  A derivation of the approximate fields to order $1/c^2$ is given by L. Page and N. I. Adams, \textit{Electrodynamics}, (Van Nos
trand, New York, 1940) pp. 171-175.

\bibitem{ } E. F. Taylor and J. A. Wheeler, \textit{Spacetime Physics} (Freeman, San Francisco, 1966), pp. 147-148, and (answers) p. 38.

\bibitem{ } The transfer of energy within the atom itself for non-circular orbits leads to the relativisitic precession of the nonrelativistic elliptical orbit.

\bibitem{ } Although the center of mass of the $e_be_{b'}$-system does not change with the orbit distortion, the center of energy of the orbit does shift in order $1/c^2$ because the average kinetic and potential energy distribution shifts.

\bibitem{ } H.A. Haus and P. Penfield Jr, Ref. 20.

\bibitem{ }  Y. Aharonov, P. Pearle, and L. Vaidman, Ref. 5 and L. Vaidman, Ref. 4.

\bibitem{ } See Y. Aharonov, P. Pearle, and L. Vaidman, Ref. 5, and L. Vaidman, Ref. 4, and also, T. H. Boyer, Ref. 4.

\bibitem{ } It is my opinion that one of the significant failures of twentieth century theoretical physics was the inconsistent application of relativistic principles. Although the Aharonov-Bohm \textit{phase shift} is an experimentally observed aspe
ct of nature, the theoretical, quantum-topological Aharonov-Bohm \textit{effect} may well go the same way as the theory of phlogiston.

\bibitem{ } T.H. Boyer, "Does the Aharonov-Bohm Effect Exist?" Found. Phys. {\bf 30}, 893-905 (2000); "Classical Electromagnetism and the Aharonov-Bohm Phase Shift," Found. Phys. {\bf 30}, 907-932 (2000).

\bibitem{ } T.H. Boyer, "The Aharonov-Bohm Effect as a Classical Electromagntic-Lag Effect: an Electrostatic Analogue and Possible Experimental Test," Il Nuovo Cimento {\bf 100B}, 685-701 (1987).

\bibitem{10} T.H. Boyer, "Classical Electromagnetic Interaction of a Charged particle with a Constant-Current Solenoid," Phys. Rev. D {\bf 8}, 1667-1679 (1973); see also, T.H. Boyer, Refs. 29, 40, 41.

\bibitem{ } Refs. 40 and 41 were published elsewhere after being rejected by  Physical Review A.

\bibitem{ } In his review of two recent manuscripts, (Ref. 40), rejected by The Physical Review,  the referee wrote the following regarding the first paper:  "The central concrete point of the paper, which has been published many times before by the
same author, is that there is a non-vanishing classical force on a charge passing the solenoid in the usual AB effect.  That assertion is false.  A charge moving in a field-free region experiences no force.  If an electromagnetic field is induced by
the passage of the charge, the resulting force on the charge must be proportional to $q^2$.  The author states that it is proportional to $(v/c)^2$, but neglects to say that the same argument makes it proportional to $(qv/c)^2$.  In fact, the force i
s exactly zero in the AB case."

\bibitem{23} T.H. Boyer, "Proposed Aharonov-Casher effect: Another example of an Aharonov-Bohm effect arising from a classical lag," Phys. Rev. A {\bf 36}, 5083-5086 (1987)and also T. H. Boyer, Ref. 4.

\bibitem{ } Y. Aharonov, P. Peale, and L. Vaidman, Ref. 5.

\bibitem{ } D. J. Griffiths, "Dipoles at rest," Am. J. Phys. {\bf 60}, 979-987 (1992), ft. nt. 31.

\bibitem{ } The analysis of the present article presents further justification for the third-law proposal used in Ref. 40.

\bibitem{ } See, for example, E. Nelson, "Derivation of the Schroedinger Equation from Newtonian Mechanics,"   Phys. Rev. {\bf 150}, 1079-1085 (1966).

\bibitem{23} See, for example, T.W. Marshall, "Statistical electrodynamics," Proc. Camb. Phil. Soc. {\bf 61}, 537-546 (1965); T.H. Boyer, "Random electrodynamics: The theory of classical electrodynamics with classical electromagntic zero-point radiat
ion," Phys. Rev. D {\bf 11}, 790-808 (1975); D. C. Cole, "Reinvestigation of the thermodynamics of blackbody radiation via classical physics," Phys. Rev. A {\bf 45}, 8471-8489 (1992).

\bibitem{ } D. J. Griffiths, \textit{Introduction to Quantum Mechanics}, (Prentice Hall, Englewood Cliffs, NJ, 1995), p. 349.

\bibitem{ } See Tonomura, (Ref. 7), p. 117; B. Lischke, "Direckte Beobachtung einzelner magnetischer Fluss quanten in supraleitenden Hohlzylindern. III," Z. Physik {\bf 239}, 360-378 (1970).  One referee for my manuscripts in Ref. 40 was unaware that
 an Aharonov-Bohm phase shift of half a fringe can be seen for a superconducting solenoid or toroid.

\bibitem{23} T.H. Boyer, "Penetration of the electric and magnetic velocity fields of a nonrelativistic point charge into a conducting plane," Phys. Rev. A {\bf 36}, 68-82 (1974); W. H. Furry, "Shielding of the Magnetic Field of a Slowly Moving Point
 Charge by a Conducting Surface," Am. J. Phys. {\bf 42}, 649-667 (1974); T. H. Boyer, "Penetration of electromagnetic velocity fields through a conducting wall of finite thickness," Phys. Rev. A {\bf 53}, 6450-6459 (1996).

\bibitem{ } See J.D. Jackson, Ref. 11, p. 337.

\bibitem{ } D. S. Jones, "The penetration into conductors of magnetic fields from moving charges," J. Phys. A {\bf} 8, 742-750 (1975).

\bibitem{ } T.H. Boyer, "Understanding the penetration of electromagnetic velocity fields into conductors," Am. J. Phys. {\bf 67}, 954-958 (1999).

\end{references}
\end{document}